\documentstyle[preprint,aps,eqsecnum,tighten]{revtex}
\def\be{\nopagebreak[3]\begin{equation}}
\def\ee{\end{equation}}
\def\ba{\nopagebreak[3]\begin{eqnarray}}
\def\ea{\end{eqnarray}}

\def\triangle{\bigtriangleup}

\def\ov{\overline}

\def\Ab{\bar{A}}

\def\E{{\hat{E}}}
\def\Pl{\ell_P}
\def\Jd{{J^{(d)}_{S,v}}}
\def\Ju{{J^{(u)}_{S,v}}}
\def\Jt{{J^{(t)}_{S,v}}}
\def\Jdu{{J^{(d+u)}_{S,v}}}
\def\jd{j^{(d)}}
\def\ju{j^{(u)}}
\def\jdu{j^{(d+u)}}

\def\C*{$C^{\star}$}
\def\Cyl{{\rm Cyl}}

\def\ge{\geq}

\def\a{\alpha}

\def\e{\epsilon}
\def\g{\gamma}

\def\S{\Sigma}

\def\A{{\cal A}}
\def\ab{\ov{\cal A}}
\def\ag{{{\cal A}/{\cal G}}}
\def\agb{{\overline {{\cal A}/{\cal G}}}}

\def\C{{\cal C}}
\def\G{{\cal G}}
\def\Gb{{\overline \G}}
\def\H{{\cal H}}

\def\Ho{{\H}^o}
\def\N{{\cal N}}

\def\Comp{{\mathchoice
{\setbox0=\hbox{$\displaystyle\rm C$}\hbox{\hbox to0pt
{\kern0.4\wd0\vrule height0.9\ht0\hss}\box0}}
{\setbox0=\hbox{$\textstyle\rm C$}\hbox{\hbox to0pt
{\kern0.4\wd0\vrule height0.9\ht0\hss}\box0}}
{\setbox0=\hbox{$\scriptstyle\rm C$}\hbox{\hbox to0pt
{\kern0.4\wd0\vrule height0.9\ht0\hss}\box0}}
{\setbox0=\hbox{$\scriptscriptstyle\rm C$}\hbox{\hbox to0pt
{\kern0.4\wd0\vrule height0.9\ht0\hss}\box0}}}}
\def\Co{{\mathchoice
{\setbox0=\hbox{$\displaystyle\rm C$}\hbox{\hbox to0pt
{\kern0.4\wd0\vrule height0.9\ht0\hss}\box0}}
{\setbox0=\hbox{$\textstyle\rm C$}\hbox{\hbox to0pt
{\kern0.4\wd0\vrule height0.9\ht0\hss}\box0}}
{\setbox0=\hbox{$\scriptstyle\rm C$}\hbox{\hbox to0pt
{\kern0.4\wd0\vrule height0.9\ht0\hss}\box0}}
{\setbox0=\hbox{$\scriptscriptstyle\rm C$}\hbox{\hbox to0pt
{\kern0.4\wd0\vrule height0.9\ht0\hss}\box0}}}}
\def\Rl{{\mathchoice
{\setbox0=\hbox{$\displaystyle\rm R$}\hbox{\hbox to0pt
{\kern0.4\wd0\vrule height0.9\ht0\hss}\box0}}
{\setbox0=\hbox{$\textstyle\rm R$}\hbox{\hbox to0pt
{\kern0.4\wd0\vrule height0.9\ht0\hss}\box0}}
{\setbox0=\hbox{$\scriptstyle\rm R$}\hbox{\hbox to0pt
{\kern0.4\wd0\vrule height0.9\ht0\hss}\box0}}
{\setbox0=\hbox{$\scriptscriptstyle\rm R$}\hbox{\hbox to0pt
{\kern0.4\wd0\vrule height0.9\ht0\hss}\box0}}}}

\def\vpi{\vec{\pi}}
\def\c{\vec{c}}
\def\vrho{\vec{\rho}}
\def\M{\vec{M}}

\begin{document}

\title{Quantum Theory of Geometry I: Area Operators\footnote{It is a
pleasure to dedicate this article to Professor Andrzej Trautman, who
was one of the first to recognize the deep relation between geometry
and the physics of gauge fields \cite{TR1,TR2} which lies at the heart
of this investigation.}}

\author{Abhay Ashtekar${}^{1}$\thanks{Electronic address: 
ashtekar@phys.psu.edu}
and Jerzy Lewandowski${}^{2,3}$\thanks{Electronic address: 
jerzy.lewandowski@fuw.edu.pl}}
\address{${}^1$Center for Gravitational Physics and Geometry\\
Physics Department, Penn State, University Park, PA 16802, USA} 
\address{${}^2$Institute of Theoretical Physics,\\
Warsaw University, ul Hoza 69, 00-681 Warsaw, Poland}
\address{${}^3$Max Planck Institut f\"ur Gravitationphysik,\\
Schlaatzweg 1, 14473 Potsdam, Germany}

\maketitle

\begin{abstract}

A new functional calculus, developed recently for a fully
non-perturbative treatment of quantum gravity, is used to begin a
systematic construction of a quantum theory of geometry. Regulated
operators corresponding to areas of 2-surfaces are introduced and
shown to be self-adjoint on the underlying (kinematical) Hilbert space
of states. It is shown that their spectra are {\it purely} discrete
indicating that the underlying quantum geometry is far from what the
continuum picture might suggest. Indeed, the fundamental excitations
of quantum geometry are 1-dimensional, rather like polymers, and the
3-dimensional continuum geometry emerges only on coarse graining. The
full Hilbert space admits an orthonormal decomposition into finite
dimensional sub-spaces which can be interpreted as the spaces of
states of spin systems. Using this property, the complete spectrum of
the area operators is evaluated. The general framework constructed
here will be used in a subsequent paper to discuss 3-dimensional
geometric operators, e.g., the ones corresponding to volumes of
regions.

\end{abstract}

\newpage
\begin{section} {Introduction}
\label{s1}

In his celebrated inaugural address, Riemann suggested \cite{Rie} that
geometry of space may be more than just a fiducial, mathematical
entity serving as a passive stage for physical phenomena, and may in
fact have a direct physical meaning in its own right. General
relativity proved this vision to be correct: Einstein's equations put
geometry on the same footing as matter. Now, the physics of this
century has shown us that matter has constituents and the
3-dimensional objects we perceive as solids in fact have a discrete
underlying structure. The continuum description of matter is an
approximation which succeeds brilliantly in the macroscopic regime but
fails hopelessly at the atomic scale. It is therefore natural to ask
if the same is true of geometry. Does geometry also have constituents
at the Planck scale?  What are its atoms? Its elementary excitations?
Is the space-time continuum only a ``coarse-grained'' approximation?
If so, what is the nature of the underlying quantum geometry?

To probe such issues, it is natural to look for hints in the
procedures that have been successful in describing matter. Let us
begin by asking what we mean by quantization of physical quantities.
Let us take a simple example --the hydrogen atom. In this case, the
answer is clear: while the basic observables --energy and angular
momentum-- take on a continuous range of values classically, in
quantum mechanics their spectra are discrete. So, we can ask if the
same is true of geometry. Classical geometrical observables such as
areas of surfaces and volumes of regions can take on continuous values
on the phase space of general relativity. Are the spectra of
corresponding quantum operators discrete? If so, we would say that
geometry is quantized.

Thus, it is rather easy to pose the basic questions in a precise
fashion. Indeed, they could have been formulated soon after the advent
of quantum mechanics. Answering them, on the other hand, has proved to
be surprisingly difficult. The main reason, it seems, is the
inadequacy of the standard techniques. More precisely, the traditional
approach to quantum field theory has been perturbative, where one {\it
begins} with a continuum, background geometry. It is then difficult to
see how discreteness would arise in the spectra of geometric
operators. To analyze such issues, one needs a fully non-perturbative
approach: geometric operators have to be constructed ab-initio without
assuming {\it any} background geometry. To probe the nature of quantum
geometry, we can not begin by {\it assuming} the validity of the
continuum picture. We must let quantum gravity itself decide whether
this picture is adequate at the Planck scale; the theory itself should
lead us to the correct microscopic picture of geometry.

In this paper, we will use the non-perturbative, canonical approach to
quantum gravity based on connections to probe these issues.  Over the
past three years, this approach has been put on a firm mathematical
footing through the development of a new functional calculus on the
space of gauge equivalent connections [4-11]. This calculus does not
use any background fields (such as a metric) and is therefore
well-suited to a fully non-perturbative treatment. The purpose of this
paper is to use this framework to explore the nature of quantum
geometry.

In section 2, we recall the relevant results from the new functional
calculus and outline the general strategy. In section 3, we present a
regularization of the area operator. Its properties are discussed in
section 4; in particular, we exhibit its entire spectrum.  Our
analysis is carried out in the ``connection representation'' and the
discussion is self-contained. However, at a non-technical level, there
is a close similarity between the basic ideas used here and those used
in discussions based on the ``loop representation''
\cite{9,10}. Indeed, the development of the functional calculus which
underlies this analysis itself was motivated, in a large measure, by
the pioneering work on loop representation by Rovelli and Smolin
\cite{11}. The relation between various approaches will discussed in
section 5.

The main result of this paper should have ramifications on the
statistical mechanical origin of the entropy of black holes along the
lines of \cite{BM,Car}. This issue is being investigated.
\end{section}

\begin{section} {Preliminaries} 
\label{s2}

This section is divided into three parts. In the first, we will recall
\cite{1,2} the basic structure of the quantum configuration space and,
in the second, that of the Hilbert space of (kinematic) quantum states
\cite{7}. The overall strategy will be summarized in the third part.

\begin{subsection} {Quantum configuration space}
\label{s2.1}

In general relativity, one can regard the space $\ag$ of $SU(2)$
connections modulo gauge transformations on a (``spatial'') 3-manifold
$\S$ as the classical configuration space \cite{12,13,14}. For systems
with only a finite number of degrees of freedom, the classical
configuration space also serves as the domain space of quantum wave
functions, i.e., as the quantum configuration space. For systems with
an infinite number of degrees of freedom, on the other hand, this is
not true: generically, the quantum configuration space is an
enlargement of the classical. In free field theory in Minkowski space
(as well as exactly solvable models in low space-time dimensions), for
example, while the classical configuration space can be built from
suitably smooth fields, the quantum configuration space includes all
(tempered) distributions. This is an important point because,
typically, the classical configuration spaces are of zero measure;
wave functions with support only on smooth configurations have zero
norm! The overall situation is the same in general relativity. The
quantum configuration space $\agb$ is a certain completion of $\ag$ 
\cite{1,2}.

The space $\agb$ inherits the quotient structure of $\ag$, i.e.,
$\agb$ is the quotient of the space $\ab$ of generalized connections
by the space $\Gb$ of generalized gauge transformations. To see the
nature of the generalization involved, recall first that each smooth
connection defines a holonomy along paths
\footnote{For technical reasons, we will assume that all paths are
analytic. An extension of the framework to allow for smooth paths is
being carried out \cite{15}. The general expectation is that the main
results will admit natural generalizations to the smooth category. In
this article, $A$ has the physical dimensions of a connection, $({\rm
length})^{-1}$ and is thus related to the configuration variable
$A_{\rm old}$ in the literature by $A = GA_{\rm old}$ where $G$ is
Newton's constant.} %
in $\S$: $h_p(A):= {\cal P}\exp -\int_p A$. Generalized connections
capture this notion. That is, each $\Ab$ in $\ab$ can be defined
\cite{3,5} as a map which assigns to each oriented path $p$ in $\S$ an
element $\Ab(p)$ of $SU(2)$ such that: i) $\Ab(p^{-1}) =
(\Ab(p))^{-1}$; and, ii) $\Ab(p_2\circ p_1) = \Ab(p_2)\cdot\Ab(p_1)$,
where $p^{-1}$ is obtained from $p$ by simply reversing the
orientation, $p_2\circ p_1$ denotes the composition of the two paths
(obtained by connecting the end of $p_1$ with the beginning of $p_2$)
and $\Ab(p_2)\cdot\Ab(p_1)$ is the composition in $SU(2)$. A
generalized gauge transformation is a map $g$ which assigns to each
point $v$ of $\S$ an $SU(2)$ element $g(x)$ (in an arbitrary, possibly
discontinuous fashion). It acts on $\Ab$ in the expected manner, at
the end points of paths: $\Ab(p)\rightarrow g(v_+)^{-1}
\cdot\Ab(p)\cdot g(v_-)$, where $v_{-}$ and $v_+$ are respectively the
beginning and the end point of $p$. If $\Ab$ happens to be a smooth
connections, say $A$, we have $\Ab(p) = h_p(A)$.  However, in general,
$\Ab(p)$ can not be expressed as a path ordered exponential of a
smooth 1-form with values in the Lie algebra of $SU(2)$
\cite{2}. Similarly, in general, a generalized gauge transformation
can not be represented by a smooth group valued function on $\S$.

At first sight the spaces $\ab$, $\Gb$ and $\agb$ seem too large to be
mathematically controllable. However, they admit three
characterizations which enables one to introduce differential and
integral calculus on them \cite{1,2,4}. We will conclude this
sub-section by summarizing the characterization --as suitable limits
of the corresponding spaces in lattice gauge theory-- which will be
most useful for the main body of this paper.

We begin with some definitions.  An {\it edge} is an oriented,
1-dimensional sub-manifold of $\S$ with two boundary points, called
{\it vertices}, which is analytic everywhere, including the vertices.
A {\it graph} in $\S$ is a collection of edges such that if two
distinct edges meet, they do so only at vertices.  In the physics
terminology, one can think of a graph as a ``floating lattice'', i.e.,
a lattice whose edges are not required to be rectangular. (Indeed,
they may even be non-trivially knotted!) Using the standard ideas from
lattice gauge theory, we can construct the configuration space
associated with the graph $\g$. Thus, we have the space $\A_\g$, each
element $A_\g$ of which assigns to every edge in $\g$ an element of
$SU(2)$ and the space $\G_\g$ each element $g_\g$ of which assigns to
each vertex in $\g$ an element of $SU(2)$. (Thus, if $N$ is the number
of edges in $\g$ and $V$ the number of vertices, $\A_\g$ is isomorphic
with $[SU(2)]^N$ and $\G_\g$ with $[SU(2)]^V$). $\G_\g$ has the
obvious action on $\A_\g$: $A_\g(e) \rightarrow g(v_+)^{-1}\cdot
A_\g(e)\cdot g(v_-)$.  The (gauge invariant) configuration space
associated with the floating lattice $\gamma$ is just
$\A_\g/\G_\g$. The spaces $\ab$, $\Gb$ and $\agb$ can be obtained as
well-defined (projective) limits of the spaces $\A_\g$, $\G_\g$ and
$\A_\g/\G_\g$ \cite{4,2}. Note however that this limit is {\it not}
the usual ``continuum limit'' of a lattice gauge theory in which one
lets the edge length go to zero. Here, we are already in the continuum
and have available to us all possible floating lattices from the
beginning. We are just expressing the quantum configuration space of
the continuum theory as a suitable limit of the configuration spaces
of theories associated with all these lattices.

To summarize, the quantum configuration space $\agb$ is a specific
extension of the classical configuration space $\ag$. Quantum states
can be expressed as complex-valued, square-integrable functions on
$\agb$, or, equivalently, as $\Gb$-invariant square-integrable
functions on $\ab$.  As in Minkowskian field theories, while $\ag$ is
dense in $\agb$ topologically, measure theoretically it is generally
sparse; typically, $\ag$ is contained in a subset set of zero measure
of $\agb$ \cite{4}.  Consequently, what matters is the value of wave
functions on ``genuinely '' generalized connections.  In contrast with
the usual Minkowskian situation, however, $\ab$, $\Gb$ and $\agb$ are
all {\it compact} spaces in their natural (Gel'fand) topologies
[4-8]. This fact simplifies a number of technical issues.

Our construction can be compared with the general framework of
`second quantization' proposed by Kijowski \cite{Kij} already twenty
years ago. He introduced the space of states for a field theory by
using the projective limit of spaces of states associated to a family
of finite dimensional theories. He also, suggested, as an example, the
lattice approach. The common element with the present approach is that
in our case the space of measures on $\ab$ is also the projective
limit of the spaces of measures defined on finite dimensional spaces
$A_\g$.

\end{subsection}

\begin{subsection} {Hilbert space}
\label{s2.2}

Since $\agb$ is compact, it admits regular (Borel, normalized)
measures and for every such measure we can construct a Hilbert space
of square-integrable functions.  Thus, to construct the Hilbert space
of quantum states, we need to select a specific measure on $\agb$.

It turns out that $\ab$ admits a measure $\mu^o$ that is preferred by
both mathematical and physical considerations \cite{2,3}.
Mathematically, the measure $\mu^o$ is natural because its definition
does not involve introduction of any additional structure: it is
induced on $\ab$ by the Haar measure on $SU(2)$. More precisely, since
$\A_\g$ is isomorphic to $[SU(2)]^N$, the Haar measure on $SU(2)$
induces on it a measure $\mu^o_\g$ in the obvious fashion.  As we vary
$\g$, we obtain a family of measures which turn out to be compatible
in an appropriate sense and therefore induce a measure $\mu^o$ on
$\ab$.  This measure has the following attractive properties \cite{2}:
i) it is faithful; i.e., for any continuous, non-negative function $f$
on $\ab$, $\int d\mu^o \, f \ge 0$, equality holding if and only if
$f$ is identically zero; and, ii) it is invariant under the (induced)
action of ${\rm Diff}[\S]$, the diffeomorphism group of $\S$.
Finally, $\mu^o$ induces a natural measure $\tilde{\mu}^o$ on $\agb$:
$\tilde{\mu}^o$ is simply the push-forward of $\mu^o$ under the
projection map that sends $\ab$ to $\agb$.  Physically, the measure
$\tilde{\mu}^o$ is selected by the so-called ``reality conditions''.
More precisely, the classical phase space admits an (over)complete set
of naturally defined configuration and momentum variables which are
real, and the requirement that the corresponding operators on the
quantum Hilbert space be self-adjoint selects for us the measure
$\tilde{\mu}^o$ \cite{7}. 

Thus, it is natural to use $\tilde\Ho := L^2(\agb, d\tilde{\mu}^o)$ as
our Hilbert space. Elements of $\tilde\Ho$ are the kinematic states;
we are yet to impose quantum constraints. Thus, $\tilde\Ho$ is the
classical analog of the {\it full} phase-space of quantum gravity
(prior to the introduction of the constraint sub-manifold). Note that
these quantum states can be regarded also as {\it gauge invariant}
functions on $\ab$. In fact, since the spaces under consideration are
compact and measures normalized, we can regard $\tilde\Ho$ as the
gauge invariant {\it sub-space} of the Hilbert space $\Ho := L^2(\ab,
d\mu^o)$ of square-integrable functions on $\ab$ \cite{3,4}. {\em In
what follows, we we will often do so}.

What do ``typical'' quantum states look like? To provide an intuitive
picture, we can proceed as follows. Fix a graph $\g$ with $N$ edges and
consider functions $\Psi_\g$ of generalized connections of the form
$\Psi_\g (\Ab) = \psi (\Ab(e_1),..., \Ab(e_N))$ for {\it some} smooth
function $\psi$ on $[SU(2)]^N$, where $e_1, ..., e_N$ are the edges of
the graph $\g$.  Thus, the functions $\Psi_\g$ know about what the
generalized connections do only to those paths which constitute the
edges of the graph $\g$; they are precisely the quantum states of the
gauge theory associated with the ``floating lattice'' $\g$. This space
of states, although infinite dimensional, is quite ``small'' in the
sense that it corresponds to the Hilbert space associated with a
system with only a {\it finite} number of degrees of freedom. However,
if we vary $\g$ through all possible graphs, the collection of all
states that results is very large.  Indeed, one can show that it is
{\it dense} in the Hilbert space $\Ho$. (If we restrict ourselves to
$\Psi_\g$ which are gauge invariant, we obtain a dense sub-space in
$\tilde\Ho$.) Since each of these states depends only on a finite
number of variables, borrowing the terminology from the quantum theory
of free fields in Minkowski space, they are called {\it cylindrical
functions} and denoted by $\Cyl$. Gauge invariant cylindrical
functions represent the ``typical'' kinematic states.  In many ways,
$\Cyl$ is analogous to the space $C_o^\infty(R^3)$ of smooth functions
of compact support on $R^3$ which is dense in the Hilbert space
$L^2(R^3, d^3x)$ of quantum mechanics. Just as one often defines
quantum operators -- e.g., the position, the momentum and the
Hamiltonians-- on $C^\infty_o$ first and then extends them to an
appropriately larger domain in the Hilbert space $L^2(R^3, d^3x)$, we
will define our operators first on $\Cyl$ and then extend them
appropriately.

Cylindrical functions provide considerable intuition about the nature
of quantum states we are led to consider. These states represent
1-dimensional polymer-like excitations of geometry/gravity rather than
3-dimensional wavy undulations on flat space. Just as a polymer,
although intrinsically 1-dimensional, exhibits 3-dimensional
properties in sufficiently complex and densely packed configurations,
the fundamental 1-dimensional excitations of geometry can be packed
appropriately to provide a geometry which, when coarse-grained on scales
much larger than the Planck length, lead us to continuum geometries
\cite{9,18}.  Thus, in this description, gravitons can arise only as
approximate notions in the low energy regime \cite{19}.  At the basic
level, states in $\tilde\Ho$ are fundamentally different from the Fock
states of Minkowskian quantum field theories. The main reason is the
underlying diffeomorphism invariance: In absence of a background
geometry, it is not possible to introduce the familiar Gaussian
measures and associated Fock spaces.

\end{subsection}

\begin{subsection}{Statement of the problem}
\label{s2.3}

We can now outline the general strategy that will be followed in
sections 4 and 5.

Recall that the classical configuration variable is an $SU(2)$
connection%
\footnote{We assume that the underlying 3-manifold $\S$ is orientable.
Hence, principal $SU(2)$ bundles over $\S$ are all topologically
trivial. Therefore, we can represent the $SU(2)$ connections on the
bundle by a $su(2)$-valued 1-form on $\S$. The matrices $\tau_i$ are
anti-Hermitian, given, e.g., by $(-i/2)$-times the Pauli matrices.} %
$A_a^i$ on a 3-manifold $\S$, where $i$ is the $su(2)$-internal index
with respect to a basis $\tau_i$. Its conjugate momentum $E^b_j$ has
the geometrical interpretation of an orthonormal triad with density
weight one \cite{20,12}, the precise Poisson brackets being:
\be
\{A_a^i(x), E^b_j(y)\} = G \delta_a^b \delta^i_j \delta^3(x,y)\, ,
\ee
where $G$ is Newton's constant. (Recall from footnote 1 that the field
$A$, used here, is related to $A_{\rm old}$ used in the literature
\cite{22} via $A= GA_{\rm old}$.)

Therefore, geometrical observables --functionals of the 3-metric-- can
be expressed in terms of this field $E^a_i$. Fix within the 3-manifold
$\S$ any analytic, finite 2-surface $S$ without boundary such that the
closure of $S$ in $\S$ is a compact. The area $A_S$ of $S$ is a
well-defined, real-valued function on the {\it full} phase space of
general relativity (which happens to depend only on $E^a_i$).  It is
easy to verify that these kinematical observables can be expressed as:
\be\label{2.4} 
A_S := \int_S dx^1\wedge dx^2 \, [E^3_i E^{3i}]^{1\over 2}\, ,
\ee
where, for simplicity, we have used adapted coordinates such that $S$
is given by $x^3 = 0$, and $x^1,x^2$ parameterize $S$, and where the
internal index $i$ is raised by a the inner product we use on $su(2)$,
$k(\tau_i,\tau_j) = -2{\rm Tr}(\tau_i\tau_j)$.

Our task is to find the corresponding operators on the kinematical
Hilbert space $\tilde\Ho$ and investigate their properties.

There are several factors that make this task difficult.  Intuitively,
one would expect that $E^a_i(x)$ to be replaced by the
``operator-valued distribution'' $-i\hbar G\delta/\delta A_a^i(x)$.
Unfortunately, the classical expression of $A_S$ involves {\it
square-roots of products} of $E$'s and hence the formal expression of
the corresponding operator is badly divergent.  One must introduce a
suitable regularization scheme.  Unfortunately, we do not have at our
disposal the usual machinery of Minkowskian field theories and even
the precise rules that are to underlie such a regularization are not
apriori clear.  

There are however certain basic expectations that we can use as
guidelines: i) the resulting operators should be well-defined on a
dense sub-space of $\tilde\Ho$; ii) their final expressions should be
diffeomorphism covariant, and hence, in particular, independent of any
background fields that may be used in the intermediate steps of the
regularization procedure; and, iii) since the classical observables
are real-valued, the operators should be self-adjoint. These
expectations seem to be formidable at first. Indeed, these demands are
rarely met even in Minkowskian field theories; in presence of
interactions, it is extremely difficult to establish rigorously that
physically interesting operators are well-defined and self-adjoint. As
we will see, the reason why one can succeed in the present case is
two-folds. First, the requirement of diffeomorphism covariance is a
powerful restriction that severely limits the possibilities. Second,
the background independent functional calculus is extremely
well-suited for the problem and enables one to circumvent the various
road blocks in subtle ways.

Our general strategy will be following. We will define the regulated
versions of area operators on the dense sub-space $\Cyl$ of
cylindrical functions and show that they are essentially self-adjoint
(i.e., admit unique self-adjoint extensions to $\tilde\Ho$).  This
task is further simplified because the operators leave each sub-space
$\H_\g$ spanned by cylindrical functions associated with any one graph
$\g$ invariant. This in effect reduces the field theory problem (i.e.,
one with an infinite number of degrees of freedom) to a quantum
mechanics problem (in which there are only a finite number of degrees
of freedom).  Finally, we will find that the operators in fact leave
invariant certain {\it finite} dimensional sub-space of $\Ho$
(associated with extended spin networks, introduced in Sec. 4.2).
This powerful simplification further reduces the task of investigating
the properties of these operators; in effect, the quantum mechanical
problem (in which the Hilbert space is still infinite dimensional) is
further simplified to a problem involving spin systems (where the
Hilbert space is finite dimensional).  It is because of these
simplifications that a complete analysis is possible.
\end{subsection}
\end{section}

\begin{section}{Regularization}
\label{s3}

Our task is to construct a well-defined operator $\hat{A}_S$ starting
from the classical expression (\ref{2.4}).  As is usual in quantum
field theory, we will begin with the formal expression obtained by
replacing $E^3_i$ in (\ref{2.4}) by the corresponding operator valued
distribution $\E^3_i$ and then regulate it to obtain the required
$\hat{A}_S$. (For an early discussion of non-perturbative
regularization, see, in particular, \cite{BP}). Our discussion will be
divided in to two parts. In the first, we introduce the basic tools
and, in the second, we apply them to obtain a well-defined operator
$\hat{A}_S$.

To simplify the presentation, let us first assume that $S$ is covered
by a single chart of adapted coordinates. Extension to the general
case is straightforward: one mimics the procedure used to define the
integral of a differential form over a manifold. That is, one takes
advantage of the coordinates invariance of the the resulting `local'
operator and uses a partition of unity.

\begin{subsection}{Tools}
\label{s3.1}

The regularization procedure involves two main ingredients. We will
begin by summarizing them.
                        
The first involves smearing of (the operator analog of) $E^3_i(x)$ and
point splitting of the integrand in (\ref{2.4}). Since in this
integrand, the point $x$ lies on the 2-surface $S$, let us try to use
a 2-dimensional smearing function. Let $f_\epsilon (x, y)$ be a
1-parameter family of fields on $S$ which tend to the $\delta(x,y)$ as
$\epsilon$ tends to zero; i.e., such that
\be \lim_{\epsilon\rightarrow 0} \int_S d^2y \, 
f_\epsilon(x^1,x^2; y^1,y^2) g(y^1,y^2) = g(x^1,x^2)\, , 
\ee
for all smooth densities $g$ of weight $1$ and of compact support on
$S$. (Thus, $f_\e(x,y)$ is a density of weight 1 in $x$ and a function
in $y$.) The smeared version of $E^3_i(x)$ will be defined to be:
\be
[E^3_i]_f (x):=  \int_S d^2y\, f_\epsilon(x,y) E^3_i(y)\, ,
\ee
so that, as $\epsilon$ tends to zero, $[E^3_i]_f$ tends to
$E^3_i(x)$. The point-splitting strategy now provides a `regularized
expression' of area:
\ba\label{ra} 
[A_S]_f &:=& \int_S d^2x \, \big[\int_S d^2y\,
f_\e (x,y)  E^3_i(y)\, \int_S d^2z \, f_\e (x,z) 
E^{3i}(z)\, \big]^{1\over 2}\nonumber \\ 
&=& \int_S d^2x\, \big[[E^3_i]_f (x)
[E^{3i}]_f (x) \big]^{1\over 2}\, , 
\ea
which will serve as the point of departure in the next subsection.  To
simplify technicalities, we will assume that the smearing field
$f_\e(x,y)$ has the following additional properties for sufficiently
small $\e > 0$: i) for any given $y$, $f_\e (x,y)$ has compact support
in $x$ which shrinks uniformly to $y$; and, ii) $f_\e(x,y)$ is
non-negative. These conditions are very mild and we are thus left with
a large class of regulators.%
\footnote{For example, $f_\e(x,y)$ can be constructed as follows. Take
{\it any} non-negative function $f$ of compact support on $S$ such
that $\int d^2x f(x) = 1$ and set $f_\e(x,y) = (1/\e^2)f((x-y)/\e)$.
Here, we have implicitly used the given chart to give $f_\e(x,y)$ a
density weight in $x$.}

We now introduce the second ingredient. To go over to the quantum
theory, we want to replace $E^3_i$ in (\ref{ra}) by $\E^3_i =
-iG\hbar\delta/\delta A_3^i$. However, it is not apriori clear that,
even after smearing, $[\E^3_i]_f$ is a well-defined operator because:
i)our wave functions $\Psi$ are functionals of {\it generalized}
connections $\Ab$, whence it is not obvious what the functional
derivative means; and, ii) we have smeared the operator only along two
dimensions. Let us discuss these points one by one.

First, let us fix a graph $\g$ and consider a cylindrical function
$\Psi_\g$ on $\ab$,
\be\label{cyl} 
\Psi_\g(\Ab) = \psi (\Ab(e_1), .., \Ab(e_N))\, , 
\ee
where, as before, $N$ is the total number of edges of $\g$ and where
$\psi$ is a smooth function on $[SU(2)]^N$. Now, a key fact
about generalized connections is that, for any given graph
$\g$, each $\Ab$ is equivalent to some smooth connection $A$ \cite{2}:
Given any $\Ab$, there exists an $A$ such that
\be
\Ab(e_k) = h_k[A] := {\cal P}\exp\, -\!\int_{e_k} A\, ,
\ee
for all $k=1,...,N$. (For any given $\Ab$, the smooth connection $A$ is
of course not unique. However, this ambiguity does not affect the
considerations that follow.)  Hence, there is a 1-1 correspondence
between the cylindrical function $\Psi_\g$ on $\ab$ and function
$\psi(h_1(A), ..., \- h_E(A))$ on the space $\A$ of smooth connections
and we can apply the operator $[\E^3_i]_f$ to the latter. The result
is:
\ba\label{re} 
[\E^3_i]_f(x) \,\cdot\, \Psi_\g(\Ab) &=& -iG\hbar \sum_{I=1}^{N}\,
\int_S d^2y f_\e(x,y)\,\big({\delta h_I\over \delta A_a^i(y)}\big)
\!\mid_{y^3= 0}\, \big({\partial\psi\over\partial h_I}\big)(A)
\nonumber \\ &=&i\Pl^2 \int_S d^2y\, f_\e(x,y) \sum_{I=1}^{N} \big[
\int_{0}^{1} dt\, \dot{e_I}^3(t)\,
\delta({y}^1,e_I^1(t))\delta({y}^2,e_I^2(t))\delta(0,e_I^3(t))
\nonumber\\
& &\quad\quad \big(h_I(1,t)\tau^i h_I(t,0)\big)^A_B\big]
\,{\partial\psi\over \partial {h_{I}}^A_B} (A)\, , 
\ea
where, $\Pl = \sqrt{G\hbar}$ is the Planck length, the index $I$
labels the edges in the graph, $[0,1]\ni t\mapsto e_I(t)$ is any
parameterization of an edge $e_I$, $h_I(t',t) := {\cal P}\,\exp -
\int_t^{t'} A_a(e_I(s))\- \dot{e}_I^a(s)ds$ is the holonomy of the
connection $A$ along the edge $e_I$ from parameter value $t$ to $t'$.
Thus, the functional derivative has a well-defined action on
cylindrical functions; the first of the two problems mentioned above
has been overcome.

However, because of the presence of the delta distributions, it is
still not clear that $[\E^3_i]_f$ is a genuine operator (rather than a
distribution-valued operator). To explicitly see that it is, we need
to specify some further details. Given a graph $\g$, we can just
subdivide some of its its edges and thus obtain a graph $\g'$ which
occupies the same points in $\S$ as $\g$ but has (trivially) more
vertices and edges. Every function which is cylindrical with respect
to the ``smaller'' graph $\g$ is obviously cylindrical with respect to
the ``larger'' graph $\g'$ as well. The idea is to use this freedom to
simplify the discussion by imposing some conditions on our graph
$\g$. We will assume that: i) if an edge $e_I$ contains a segment
which lies in $S$, then it lies entirely in the closure of $S$; ii)
each isolated intersection of $\g$ with the 2-surface $S$ is a vertex
of $\g$; and, iii) each edge $e_I$ of $\g$ intersects $S$ at most
once.  (The overlapping edges are often called edges `tangential' to
$S$; they should not be confused with edges which `cross' $S$ but
whose tangent vector at the intersection point is tangent to $S$).  If
the given graph does not satisfy one or more of these conditions, we
can obtain one which does simply by sub-dividing of some of the
edges. Thus these conditions are not restrictive. They are introduced
to simplify the ``book-keeping'' in calculations.

Let us now return to (\ref{re}). If an edge $e_I$ has no point in
common with $S$, it does not contribute to the sum. If it is contained
in $S$, $\dot{e}^3_I$ vanishes identically whence its contribution
also vanishes. (For a subtlety, see the remark below Eq (\ref{X}).) We
are thus left with edges which intersect $S$ in isolated points. Let
us first consider only those edges which are `outgoing' at the
intersection. Then, at the intersection point, the value of the
parameter $t$ is zero and, for a given edge $e_I$, $\dot{e_I}^3$ is
positive (negative) if $e_I$ is directed ``upwards'' along increasing
$x^3$ (``downwards'' along decreasing $x^3$). Hence, $(\ref{re})$
becomes%
\footnote{In the first step, we have used the regularization 
$\int_0^\infty dz g(z) \delta (z) = \textstyle{1\over 2}g(0)$ which
follows if the $\delta(z)$ is obtained, in the standard fashion, as a
limit of functions which are symmetric about $0$.}: %
\ba [\E^3_i]_f (x) \,\cdot\, \Psi_\g &=& \, {i\Pl^2\over 2} \Big[\
\sum_{I=1}^{N} \big[\int_S d^2y \kappa_I f_\e(x,y)
\delta({y}^1,e_I^1(0))\delta({y}^2,e_I^2(0)) \big(
h_I\tau^i\big)^A_B\big]\, {\partial \psi\over\partial
{h_I}{}^A_B}\nonumber\\ &=& {i\Pl^2\over 2} \sum_{I=1}^{N}\kappa_I\,
f_\e (x,e_I(0))\,\, L_I^i\,\cdot\, \psi(\Ab(e_1), ... ,\Ab(e_N))\, ,
\ea
where, the constant $\kappa_I$ associated with the edge $e_I$ is given
by:
\be
\kappa_I= 
\cases{\,0,& if $e_I$ is tangential to $S$ or does not intersect 
      $S$, \cr
       +1,& if $e_I$ has an isolated intersection with $S$ and 
        lies above $S$ \cr 
      -1,& if $e_I$ has an isolated intersection with $S$ and 
       lies below $S$\cr} 
\ee
and where $L^i_I$ is the left invariant vector field in the $i$-th
internal direction on the copy of $SU(2)$ corresponding to the $I$-th
edge 
\be\label{L} L_I^i\,\cdot\, \psi(\Ab(e_1), ... ,\Ab(e_N)) =
(\Ab(e_I)\tau^i)^A_B\, {\partial\psi\over \partial (\Ab(e_I))^A_B}.
\ee 
If some of the edges are `incoming' at the intersection point, then
the final expression of $[\E^a_i]_f (x)$ can be written as:
\be 
[\E^3_i]_f (x) \,\cdot\, \Psi_\g =\ {i\Pl^2\over 2} 
\Big[\sum_{I=1}^{N} \kappa_I\, f_\e (x,v_{\alpha_I})\,\, X_I^i\Big]\,
\cdot\, \psi(\Ab(e_1), ... ,\Ab(e_N))\, ,
\ee
where $X_I^i$ is an operator assigned to a vertex $v$ and an edge
$e_I$ intersecting $v$ by the following formula
\be\label{X} 
X_I^i\,\cdot\, \psi(\Ab(e_1), ... ,\Ab(e_N)) = \cases{
(\Ab(e_I)\tau^i)^A_B\, 
{\partial\psi\over \partial (\Ab(e_I))^A_B},& when $e_I$ is outgoing\cr
-(\tau^i\Ab(e_I))^A_B\, 
{\partial\psi\over \partial (\Ab(e_I))^A_B},& when $e_I$ is incoming\cr}
\ee

{\sl Remark:} Let us briefly return to the edges which are tangential
to $S$. In this case, although $\dot{e}^3_I$ vanishes, we also have a
singular term $\delta(0,0)$ (in the $x^3$ direction) in (\ref{re}).
Hence, to recover an unambiguous answer, {\it for these edges}, we
need to smear also in the third direction using an additional
regulator, say $g_{\e'}(x^3, y^3)$. When this is done, one finds that
the contribution of the tangential edges vanishes even before removing
the regulator; as stated earlier, the tangential edges do not
contribute. We did not introduce the smearing in the third direction
right in the beginning to emphasize the point that this step is
unnecessary for the edges whose contributions survive in the end.

The right side again defines a cylindrical function based on the
(same) graph $\g$. Denote by $\Ho_\g$ the Hilbert space $L^2(\A_\g,
d\mu^o_\g)$ of square integrable cylindrical functions associated with
a fixed graph $\g$. Since $\mu^o_\g$ is the induced Haar measure on
$\A_\g$ and since the operator is just a sum of right/left invariant
vector fields, standard results in analysis imply that, with domain
${\rm Cyl}_\g^1$ of all $C^1$ cylindrical functions based on $\g$, it
is an essentially self-adjoint on $\Ho_\g$. Now, it is straightforward
to verify that the operators on $\Ho_\g$ obtained by varying $\g$ are
all compatible%
\footnote{Given two graphs, $\g$ and $\g'$, we say that $\g\ge \g'$ if
and only if every edge of $\g'$ can be written as a composition of
edges of $\g$. Given two such graphs, there is a projection map from
$\A_\g$ to $\A_{\g'}$, which, via pull-back, provides an unitary
embedding $U_{\g,\g'}$ of $\tilde\Ho_{\g'}$ into $\tilde\Ho_\g$. A
family of operators ${\cal O}_\g$ on the Hilbert spaces $\Ho_\g$ is
said to be compatible if $U_{\g,\g'}{\cal O}_{\g'} = {\cal O}_{\g}
U_{\g,\g'}$ and $U_{\g,\g'}D_{\g'}\subset D_{\g}$ for all $\g\ge \g'$,
where $D_{\g}$ and $D_{\g'}$ are the domains of ${\cal O}_{\g}$ and
${\cal O}_{g'}$.}
in the appropriate sense. Hence, it follows from the general results
in \cite{5} that $[\E^3_i]_f(x)$, with domain ${\rm Cyl}^1$ (the space
of all $C^1$ cylindrical functions), is an essentially self-adjoint
operator on $\Ho$. For notational simplicity, we will denote its
self-adjoint extension also by $[\E^3_i]_f(x)$. (The context should
make it clear whether we are referring to the essentially self-adjoint
operator or its extension.)

The fact that this operator is well-defined may seem surprising at
first sight since we have used only a 2-dimensional smearing. Recall
however that in free field theory in Minkowski space, the action of
the momentum operator on cylindrical functions is well-defined in the
same sense without any smearing at all. In our case, a 2-dimensional
smearing is needed because our states contain one --rather than three-
dimensional excitations.

\end{subsection}
\begin{subsection}{Area operators}
\label{s3.2}

Let us now turn to the integrand of the smeared area operator
(corresponding to (\ref{ra})). Denoting the determinant of the
intrinsic metric on $S$ by $g_S$, we have:
\ba\label{rg} 
[\hat{g}_S]_f (x)\,\cdot\, \Psi_\g &:=& [E^3_i]_f (x)
[E^{3i}]_f (x) \,\cdot\, \Psi_\g\nonumber\\ 
&=& -{\Pl^4\over 4}\big[ \sum_{I,J} \kappa(I,J) f_\e 
(x, v_{\alpha_I}) f_\e (x, v_{\alpha_J})\, X_I^i X_J^i\big] \,
\cdot\, \Psi_\g\, ,
\ea
where the summation goes over all the oriented pairs $(I,J)$;
$v_{\a_I}$ and $v_{\a_J}$ are the vertices at which edges $e_I$ and
$e_J$ intersect $S$; $\kappa(I,J) = \kappa_I\kappa_J$ equals $0$ if
either of the two edges $e_I$ and $e_J$ fails to intersect $S$ or lies
entirely in $S$, $+1$ if they lie on the same side of $S$, and, $-1$
if they lie on the opposite sides. (For notational simplicity, from
now on we shall not keep track of the position of the internal indices
$i$; as noted in Sec. 2.3, they are contracted using the invariant
metric on the Lie algebra $su(2)$.) The next step is to consider
vertices $v_\a$ at which $\g$ intersects $S$ and simply rewrite the
above sum by re-grouping terms by vertices. The result simplifies if
we choose $\e$ sufficiently small so that, $f_\e(x, v_{\a_I}) f_\e(x,
v_{\a_J})$ is zero unless $v_{\a_I}= v_{\a_J}$.  We then have:
\be 
[\hat{g}_S]_f (x)\,\cdot\, \Psi_\g = - {\Pl^4\over 4}\big[\sum_{\a}\, 
 (f_\e(x, v_\a))^2\, \sum_{I_\a, J_\a} \kappa(I_\a, J_\a) 
X_{I_\a}^i X_{J_\a}^i\big]\,\cdot\, \Psi_\g\, ,
\ee
where the index $\a$ labels the vertices on $S$ and $I_\a$ and $J_\a$
label the edges at the vertex $\a$.

The next step is to take the square-root of this expression. The same
reasoning that established the self-adjointness of $[\E^3_i]_f(x)$ now
implies that $[\hat{g}_S]_f(x)$ is a non-negative self-adjoint
operator and hence has a well-defined square-root which is also a
positive definite self-adjoint operator. Since we have chosen $\e$
to be sufficiently small, for any given point $x$ in $S$, $f_\e
(x,v_\a)$ is non-zero for at most one vertex $v_\a$. We can therefore
take the sum over $\a$ outside the square-root. One then obtains
\be\label{rqae} 
([\hat{g}_S]_f)^{1\over 2} (x)\,\cdot\, \Psi_\g =  
{\Pl^2\over 2}\sum_{\a}\, f_\e(x, v_\a) \big[\sum_{I_\a, J_\a} 
\kappa(I_\a, J_\a) X_{I_\a}^i X_{J_\a}^i \big]^{1\over 2}\,\cdot\, 
\Psi_\g.
\ee
Note that the operator is neatly split; the $x$-dependence all resides
in $f_\e$ and the operator within the square-root is ``internal'' in
the sense that it acts only on copies of $SU(2)$.

Finally, we can remove the regulator, i.e., take the limit as
$\epsilon$ tends to zero. By integrating both sides against test
functions on $S$ and then taking the limit, we conclude that the
following equality holds in the distributional sense:
\be\label{qae} 
\widehat{\sqrt{{g}_S}} (x)\,\cdot\, \Psi_\g = {\Pl^2\over 2}
\sum_{\a}\,
\delta^{(2)}(x, v_\a) \big[\sum_{I_\a, J_\a} \kappa(I_\a, J_\a)
X_{I_\a}^i X_{J_\a}^i \big]^{1\over 2}\cdot\, \Psi_\g.  
\ee
Hence, the regularized area operator is given by:
\be \label{AR} 
\hat{A}_S \,\cdot\, \Psi_\g = {\Pl^2\over 2} \sum_\a\,
\big[\sum_{I_\a, J_\a} \kappa(I_\a, J_\a) X_{I_\a}^i X_{J_\a}^i
\big]^{1\over 2}\,\cdot\, \Psi_\g .  
\ee
(Here, as before, $\a$ labels the vertices at which $\g$ intersects
$S$ and $I_\a$ labels the edges of $\g$ at the vertex $v_\a$.) With
${\rm Cyl}^2$ as its domain, $\hat{A}_S$ is essentially self-adjoint
on the Hilbert space $\Ho$.

Let us now remove the assumption that the surface $\S$ is covered by a
single chart of adapted coordinates. If such a global chart does not
exist, we can cover $\S$ with a family ${\cal U}$ of neighborhoods
such that for each $U\in {\cal U}$ there exists a local coordinates
system $(x^a)$ adapted to $\S$.  Let $(\varphi_U)_{U\in{\cal U}}$ be a
partition of unity associated to ${\cal U}$. We just repeat the above
regularization for a slightly modified classical surface area
functional, namely for
\be 
A_{S,U}\ :=\ \int_S dx^1\wedge dx^2 \,\varphi_U 
[E^3_i E^{3i}]^{1\over 2}\ 
\ee
which has support within a domain $U$ of an adapted chart. Thus, we
obtain the operator $\hat{A_{SU}}$. Then we just define
\be
\hat{A}_{S}\ =\ \sum_{U\in{\cal U}}\hat{A}_{S,U}.  
\ee 
The result is given again by the formula (\ref{AR}). The reason why
the functions $\varphi_U$ disappear from the result is that the
operator obtained for a single domain of an adapted chart is
insensitive on changes of this chart. This concludes our technical
discussion.

The classical expression $A_S$ of (\ref{2.4}) is a rather complicated.
It is therefore somewhat surprising that the corresponding quantum
operators can be constructed rigorously and have quite manageable
expressions.  The essential reason is the underlying diffeomorphism
invariance which severely restricts the possible operators.  Given a
surface and a graph, the only diffeomorphism invariant entities are
the intersection vertices. Thus, a diffeomorphism covariant operator
can only involve structure at these vertices. In our case, it just
acts on the copies of $SU(2)$ associated with various edges at these
vertices.

We have presented this derivation in considerable detail to spell out
all the assumptions, to bring out the generality of the procedure and
to illustrate how regularization can be carried out in a fully
non-perturbative treatment. While one is free to introduce auxiliary
structures such as preferred charts or background fields in the
intermediate steps, the final result must respect the underlying
diffeomorphism invariance of the theory. These basic ideas will be used
repeatedly for other geometric operators in the sequel to this paper.
\end{subsection}

\begin{subsection}{General properties of operators}
\label{s3.3}

1. {\em Discreteness of the spectrum:} By inspection, it follows that
the total area operator $\hat{A}_S$ leaves the sub-space of
$\Cyl^2_\g$ which is associated with any one graph $\g$ invariant and
is a self-adjoint operator on the sub-space $\Ho_\g$ of $\Ho$
corresponding to $\g$. Next, recall that $\Ho_\g = L^2(\A_\g,
d\mu^o)$, where $\A_\g$ is a compact manifold, isomorphic with
$(SU(2))^N$ where $N$ is the total number of edges in $\g$. As, we
explained below, the restriction of $\hat{A}_S$ to $\Ho_\g$ is given
by certain commuting elliptic differential operators on this compact
manifold. Therefore, all its eigenvalues are discrete. Now suppose
that the complete spectrum of $\hat{A}_S$ on $\Ho$ has a continuous
part. Denote by $P_c$ the associated projector. Then, given any $\Psi$
in $\Ho$, $P_c\cdot\Psi$ is orthogonal to $\Ho_\g$ for any graph $\g$,
and hence to the space $\Cyl$ of cylindrical functions. Now, since
$\Cyl^2$ is dense in $\Ho$, $P_c\cdot\Psi$ must vanish for all $\Psi$
in $\Ho$. Hence, the spectrum of $\hat{A}_S$ has no continuous part.

Note that this method is rather general: It can be used to show that
{\it any} self-adjoint operator on $\Ho$ which maps (the intersection
of its domain with) $\Ho_\g$ to $\Ho_\g$, and whose action on $\Ho_\g$
is given by elliptic differential operators, has a purely discrete
spectrum on $\Ho$. Geometrical operators, constructed purely from the
triad field tend to satisfy these properties.

2. {\em Area element:} Note that not only is the total area operator
well-defined, but in fact it arises from a local area element,
$\widehat{\sqrt{g_S}}$, which is an operator-valued distribution in
the usual sense. Thus, if we integrate it against test functions, the
operator is densely defined on $\Ho$ (with $C^2$ cylindrical functions
as domain) and the matrix elements
\be 
\langle {\Psi'}_{\g'},\, \widehat{\sqrt{g_S}}(x) \,\cdot\, 
\Psi_\g\rangle \nonumber
\ee
are 2-dimensional distributions on $S$. Furthermore, since we did not
have to renormalize the regularized operator (\ref{rqae}) before
removing the regulator, there are {\it no} free renormalization
constants involved. The local operator is completely unambiguous. 

3. {\em $[\hat{g}_S]_f$ versus its square-root:} Although the
regulated operator $[\hat{g}_s]_f$ is well-defined, if we let
$\epsilon$ to go zero, the resulting operator is in fact divergent:
roughly, it would lead to the square of the 2-dimensional $\delta$
distribution. Thus, the determinant of the 2-metric is not a
well-defined in the quantum theory. As we saw, however, the
square-root of the determinant {\em is well} defined: We have to first
take the square-root of the {\em regulated} expression and {\it then}
remove the regulator. This, in effect, is the essence of the
regularization procedure.

To get around this divergence of $\hat{g}_S$, as is common in
Minkowskian field theories, we could have first rescaled
$[\hat{g}_S]_f]$ by an appropriate factor and then taken the
limit. Then result can be a well-defined operator, but it will depend
on the choice of the regulator, i.e., additional structure introduced
in the procedure. Indeed, if the resulting operator is to have the
same density character as its classical analog $g_S(x)$ --which is a
scalar density of weight two-- then the operator can not respect the
underlying diffeomorphism invariance.%
\footnote{If, on the other hand, for some reason, we are willing to
allow the limiting operator to have a {\it different} density
character than its classical analog, one can renormalize
$[\hat{g}]_f(x)$ in such a way as to obtain a background independent
limit. For instance, we may use $f_\e = (1/\e^2)
\theta(|x-x'|-{\e\over 2})$, and rescale $[\hat{g}]_f$ by $\e^2$
before taking the limit. Then the limit is a well defined,
diffeomorphism covariant operator but it is a scalar density of weight
{\it one} rather than two.} %
For, there is no metric/chart independent distribution on $S$ of
density weight two. Hence, such a `renormalized' operator is not
useful to a fully non-perturbative approach. For the square-root, on
the other hand, we need a local density of weight {\it one}. And, the
2-dimensional Dirac distribution provides this; now is no apriori
obstruction for a satisfactory operator corresponding to the area
element to exist. This is an illustration of what appears to be
typical in non-perturbative approaches to quantum gravity: Either the
limit of the operator exists as the regulator is removed without the
need of renormalization or it inherits back-ground dependent
renormalization fields (rather than constants).

4. {\em Vertex operators:} As noted already, in the final expressions
of the area element and area operators, there is a clean separation
between the `$x$-dependent' and the `internal' parts.  Given a graph
$\g$, the internal part is a sum of square-roots of the operators
\be
\triangle_{S,v_\a} := \sum_{I_\a, J_\a}\,\kappa(I_\a, J_\a) 
X^i_{I_\a} X^i_{J_\a}
\ee
associated with the surface $S$ and the vertex $v_\a$ on it. It is
straightforward to check that operators corresponding to different
vertices commute. Therefore, to analyze the properties of area
operators, we can focus just on one vertex operator at a time.

Furthermore, given the surface $S$ and a point $v$ on it, we can
define an operator $\triangle_{S,v}$ on the dense sub-space ${\rm
Cyl}^2$ on $\Ho$ as follows:
\be\label{4.3} 
\triangle_{S,v}\,\cdot\, \Psi_\g := 
\cases{\sum_{I, J}\,\kappa(I,J) X^i_{I} X^i_{J}\,\cdot\,\Psi_\g, 
& if $\g$ intersects $S$ in $v$, \cr
0, & Otherwise\cr}  
\ee
where $I$ and $J$ label the edges of $\g$ which have $v$ as a vertex.
(Recall that every cylindrical function is associated with {\it some}
graph $\g$. As before, if $\g$ intersects $S$ at $v$ but $v$ is not a
vertex of $\g$, one can extend $\g$ just by adding a new vertex $v$
and orienting the edges at $v$ to outgoing.) It is straightforward to
verify that this definition is unambiguous: if a cylindrical function
can be represented in two different ways, say as $\Psi_\g$ and
$\Psi_{\g'}$, then $\triangle_{S,v} \,\cdot\, \Psi_\g$ and
$\triangle_{S,v} \,\cdot\, \Psi_{\g'}$ are two representations of the
same function on $\ab$.  There is a precise sense \cite{5} in which
$\triangle_{S,v}$ can be regarded as a Laplacian operator on
$\Ho$. The area operator is a sum over all the points $v$ of $S$ of
square-roots of Laplacians,
\be\label{A} 
\hat{A}_S\ =\ {\Pl^2\over 2}\sum_{v\in S}\sqrt{-\triangle_{S,v}}\, . 
\ee 
(Here the sum is well defined because, for any cylindrical function,
it contains only a finite number of non-zero terms, corresponding to
the isolated intersection points of the associated graph with $S$).
We will see in the next subsection that this fact is reflected in its
spectrum.
 
5. {\em Gauge invariance:} The classical area element $\sqrt{g_S}$ is
invariant under the internal rotations of triads $E^a_i$; its Poisson
bracket with the Gauss constraint functional vanishes.  This symmetry
is preserved in the quantum theory: the quantum operator
$\widehat{\sqrt{g_S}}$ commutes with the induced action of $\Gb$ on
the Hilbert space $\Ho$. Thus, $\widehat{\sqrt{g_S}}$ and the total
area operator $\hat{A}_S$ map the space of gauge invariant states to
itself; they project down to the Hilbert space $\tilde\Ho$ of
kinematic states.

Note, however, that the regulated triad operators $[\hat{E}^3_i]_f$
are {\it not} gauge invariant; they are defined only on
$\Ho$. Nonetheless, they are useful; they feature in an important way
in our regularization scheme. In the loop representation, by contrast,
one can only introduce gauge invariant operators and hence the
regulated triad operators do not exist. Furthermore, even in the
definition (\ref{ra}) of the regularized area element, one must use
holonomies to transport indices between the two points $y$ and $z$.
While this manifest gauge invariance is conceptually pleasing, in
practice it often makes the calculations in the loop representation
cumbersome; one has to keep track of these holonomy insertions in the
intermediate steps although they do not contribute to the final
result.

6. {\em Overall Factors}: The overall numerical factors in the
expressions of various operators considered above depend on two
conventions. The first is the convention noted in footnote 4 used in
the regularization procedure. Could we not have used a different
convention, setting $\int_0^\infty dz g(z)\delta(z) = c g(0)$ and
$\int^0_{-\infty} dz g(z)\delta(z) = (1-c) g(0)$ for some constant
$c\not= 1/2$? The answer is in the negative. For, in this case, the
constant $\kappa_I$ would take values:
\be
\kappa_I= 
\cases{\,0,& if $e_I$ is tangential to $S$ or does not intersect 
       $S$, \cr
       +2c,& if $e_I$ has an isolated intersection with $S$ and 
        lies above $S$ \cr 
      -2(1-c),& if $e_I$ has an isolated intersection with $S$ and 
       lies below $S$\cr} 
\ee
It then follows that, unless $c = 1/2$, the action of the area
operator $\hat{A}_S$ on a given cylindrical function would change if
we simply reverse the orientation on $S$ (keeping the orientation on
$\S$ the same). Since this is physically inadmissible, we must have $c
= 1/2$; there is really no freedom in this part of the regularization
procedure.

The second convention has to do with the overall numerical factor in
the action, which dictates the numerical coefficients in the
symplectic structure.  Here, we have adopted the convention of
\cite{22} (see chapter 9) which makes the Poisson bracket $\{A_a^i(x),
\,E^b_j(y)\} = G \delta^b_a\delta^i_j \delta(x,y)$, enabling us to
express $\E^a_i(x)$ as $-iG\hbar\delta/\delta A_a^i(x)$. (Had we
rescaled the action by $1/8\pi$ as is sometimes done, in our
expressions, Newton's constant $G$ would be replaced by $8\pi G$.)

\end{subsection}
\end{section}

\begin{section}{Eigenvalues and Eigenvectors}
\label{s4}

This section is divided into three parts. In the first, we derive the
complete spectrum of the area operators, in the second, we extend the
notion of spin networks, and in the third, we use this extension to
discuss eigenvectors.

\begin{subsection}{The complete spectrum}
\label{s4.1}

We are now ready to calculate the complete spectrum of
$\hat{A}_S$. Since $\hat{A}_S$ is a sum of square-roots of vertex
operators which all commute with one another, the task reduces to that
of finding the spectrum of each vertex operator. Furthermore, since
vertex operators map ($C^2$) cylindrical functions associated with any
one graph to $(C^0$) cylindrical functions associated with the {\it
same} graph, we can begin with an arbitrary but fixed graph
$\g$. Consider then a vertex operator $\triangle_{S,v}$ and focus on
the edges of $\g$ which intersect $S$ at $v$. Let us divide the edges
in to three categories: let $e_1, ..., e_d$ lie `below' $S$ (`down'),
$e_{d+1}, ..., e_u$ lie `above' $S$ (`up') and let $e_{u+1}, ..., e_t$
be tangential to $S$. (As before, the labels `down' and `up' do not
have an invariant significance; the orientation of $S$ and of $\S$
enable us to divide the non-tangential edges in to two parts and we
just label one as `down' and the other as `up'.) Let us set:
\ba\label{4.4} 
\Jd^i{}_\g&:= -i\, (X^i_1 + ... + X^i_d),\quad
\Ju^i{}_\g &:= -i \, (X^i_{d+1} + ... + X^i_u), \nonumber \\ 
\Jt^i{}_\g &:= -i\, (X^i_{u+1}+ ... + X^i_t), \quad
\Jdu^i{}_\g &:= \Jd^i + \Ju^i 
\ea
where $X^i_I$ is the operator defined in (\ref{X}) assigned to the
point $v$ and an edge $e_I$ at $v$.  This notation is suggestive. We
can associate with each edge $e$ a particle with only a spin degree of
freedom. Then, the operators $-i X^i_e$ can be thought of as the $i$th
component of angular momentum operators associated with that particle
and $\Jd^i, \Ju^i$ and $\Jt^i$ as the total `down', `up' and
`tangential' angular momentum operators at the vertex $v$.

By varying the graph, we thus obtain a family of operators. It is easy
to check that they satisfy the compatibility conditions and thus
define operators $\Ju^i, \Jd^i. \Jt^i$ and $\Jdu^i$ on $\Cyl$. It is
also easy to verify that they all commute with one another. Hence, one
can express the vertex operator $\triangle_{S,v}$ simply as:
\be\label{4.5}
-\triangle_{S,v} = (\Jd^i - \Ju^i)(\Jd^i - \Ju^i)\, ;
\ee
because of the factor $\kappa(I, J)$ in (\ref{4.3}), the edges which
are tangential do not feature in this expression.

The evaluation of possible eigenvalues is now straightforward. It is
simplest to express $\triangle_{S,v}$ as
\be\label{4.6} 
-\triangle_{S,v} = 2(\Jd)^2 +  2(\Ju)^2 - (\Jdu)^2\, .  
\ee 
and, as in elementary text-books, go to the representation in which
the operators $(\Jd)^2, (\Ju)^2$ and $(\Jdu)^2$ are diagonal.  If we
restrict now the operators to $\Cyl_\g$ associated to a fixed graph,
it is obvious that the possible eigenvalues $\lambda$ of
$\triangle_{S,v}$ are given by:
\be\label{4.7} 
\lambda_{S,v} = 2\jd(\jd+1) + 2\ju(\ju+1) - \jdu (\jdu+1) 
\ee
where $\jd, \ju$ and $\jdu$ are half integers subject to the usual
condition:
\be\label{4.8}
\jdu \in \{ |\jd - \ju|, |\jd - \ju|+1,\, ... \, , \jd+\ju\}.   
\ee

Returning to the total area operator, we note that the vertex
operators associated with distinct vertices commute. Although the sum
(\ref{A}) is not finite, restricted to any graph $\g$ and $\Cyl_\g$ it
becomes finite. Therefore, the eigenvalues $a_S$ of $\hat{A}_S$ are
given by
\be\label{4.9}
a_S = {\Pl^2\over 2}\, \sum_{\alpha} \Big[2\jd_\alpha(\jd_\alpha+1) + 
2\ju_\alpha(\ju_\alpha+1) 
- \jdu_\alpha (\jdu_\alpha+1)\Big]^{1\over 2}
\ee
where $\a$ labels a finite set of points in $S$ and the non-negative
half-integers assigned to each $\a$ are subject to the inequality
(\ref{4.8}).  The question now is if all these eigenvalues are
actually attained, i.e., if, given {\it any} $a_S$ of the form
(\ref{4.9}), there are eigenvectors in $\Ho$ with that eigenvalue. In
Sec. 4.3, will show that the full spectrum is indeed realized on
$\Ho$.

The area operators map the subspace $\tilde\Ho$ of {\it gauge
invariant} elements of $\Ho$ to itself. Hence we can ask for their
spectrum on $\tilde\Ho$. We will see in Sec. 4.3 that further
restrictions can now arise depending on the topology of the surface
$S$. There are three cases:
\begin{itemize}
\item{i)} The case when $S$ is an open surface whose closure is 
contained in $\Sigma$. An example is provided by the disk $z= 0,\,
x^2+y^2 < r_o$ in $R^3$.  In this case, there is no additional
condition; all $a_S$ of (\ref{4.9}) subject to (\ref{4.8}) are
realized. 

\item{ii)} The case when the surface $S$ is closed ($\partial S = 
\emptyset$) and divides $\Sigma$ into disjoint open sets
$\Sigma_1$ and $\Sigma_2$ (i.e., $\Sigma = \Sigma_1 \cup S \cup
\Sigma_2$ with $\Sigma_1\cap \Sigma_2 = \emptyset$.) An example
is given by: $\Sigma = R^3$ and $S= S^2$. In this case, there is an
a condition on the half integers $\jd_\alpha$ and $\ju_\alpha$
that appear in (\ref{4.9}) in addition to (\ref{4.8}):
\be\label{sphere}
\sum_\alpha \jd_\alpha = N, \quad {\rm and} \quad \sum_\alpha 
\ju_\alpha = N' \ee
for some integers $N$ and $N'$.
\item{iii)} The case when $S$ is closed but not of type ii). An
example is given by; $\Sigma = S^1\times S^1\times S^1$ and $S =
S^1\times S^1$. In this case, the additional condition is milder:
\be\label{torus}
\sum_\alpha (\jd_\alpha + \ju_\alpha) = N \ee
for some interger $N$.  
\end{itemize}

Next, let us note some properties of this spectrum of $\hat{A}_S$. By
inspection, it is clear that the smallest eigenvalue is $0$ and that
the spectrum is unbounded from above. One can ask for the `area gap'
i.e., the value of the smallest non-zero eigenvalue. On the full
Hilbert space $\Ho$, it is given by:
\be a_{S}^o\ =\ {\sqrt{3}\over 4} \Pl^2.  
\ee 
This is a special case of the situation when there is only one term in
the sum in (\ref{4.9}) with $\jd = 0$, $\ju=\jdu =j$. Then
\be\label{simplecase}  
a_S\ =\ {\Pl^2\over  2}\sqrt{j(j+1)}\, ,
\ee 
and, if we choose $j={1\over 2}$, we obtain the eigenvalue $a_{S}^o$.
On the Hilbert space $\tilde\Ho$ of gauge invariant states, on the
other hand, because of the constraints on the spectrum discussed
above, the area gap is sensitive to the topology of $S$:
\ba a_{S}^o & = & {\sqrt{3}\over 4} \Pl^2\,\,\, \hbox{\rm if $S$ is of
type i)} \nonumber\\ 
a_{S}^o & = & {2\sqrt{2}\over 4}\Pl^2 \,\,\,
\hbox{\rm if $S$ is of type ii)}\, \nonumber\\ 
a_{S}^o & = & {2\over4} \Pl^2\,\,\, \hbox{\rm if $S$ is of type iii)}. 
\ea

Another important feature of the spectrum is its behavior for large
$a_S$. As noted above, the spectrum is discrete. However, an
interesting question is if it approaches continuum and, if so, in what
manner. We will now show that as $a_S\rightarrow\infty$, the
difference $\Delta a_S$ between $a_S$ and its closest eigenvalue
satisfies the inequality
\be \label{ineq} 
\Delta a_S\ \le\ (\Pl^2/ 2)(\Pl/ \sqrt{a_S}) 
+ O((\Pl^2/ a_S))\Pl^2 \ee
and hence tends to zero (irrespective of the topology of
$S$). Specifically, given (odd) integers $M$ and $N$ satisfying $1\le
M\le 2\sqrt{N}$, we will obtain an eigenvalue $a_{S, N, M}$ of
$\hat{A}_S$ such that for sufficiently large $N$, the bound
(\ref{ineq}) is explicitly realized.
\footnote{This calculation was motivated by the results of Bekenstein
and Mukhanov \cite{BM} and our estimate has an interesting implication
on whether the Hawking spectrum is significantly altered due to
quantum gravity effects. Because the ``level spacing'' $\Delta a_S$
goes to zero as $a_S$ goes to infinity, the considerations of
\cite{BM} do not apply to {\it large} black holes in our approach and
there is no reason to expect deviations from Hawking's semi-classical
results. On the other hand, for small black-holes, --i.e., the final
stages of evaporation-- the estimate does not apply and one expects
transitions between area eigenstates to show significant deviations.}
Let us label representations of $SU(2)$ by their dimension, $
n_\alpha\ =\ 2j_\alpha +1$.  Let $n_\alpha$, $\alpha=1,...,M$ be (odd)
integers such that $\sum_{\alpha=1}^M\,n_\alpha\ =\ N$, and
$|n_\alpha-{N\over M}|\ <\, 2$ Then, for each $M$, we have from
(\ref{simplecase}) an eigenvalue $a_{S,N,M}$
\ba 
a_{S,N,M} \ &= &\ {\Pl^2\over 2} \sum_{\a=1}^M \sqrt{j_\a(j_a+1)}\ 
\nonumber\\ 
&=& \ {\Pl^2\over 4}\sum_{\a=1}^M \big(n_\a -{1\over 2n_\a}) + O({1\over
N})\big)\nonumber\\ 
&=& \ {\Pl^2\over 4}(N\ -\ {M^2\over 2N}\ + {kM^2\over N^2} + 
O({1\over N}))\,  
\ea 
for some integer $k\in [1,\, M/2]$.  As $M$ varies between $1$ and
$2\sqrt{N}$, $a_{S, N,M}$ varies between $(\Pl^2/4)N$ and $(\Pl^2/4)
(N-2) + 4k/N\, \le\, (\Pl^2/4)((N-2) +4/\sqrt{N})$. Hence, given a
sufficiently large $a_S$, there exist integers $N,M$ satisfying the
conditions given above such that $\Delta a_S := |a_S - a_{n,m}|$
satisfies the inequality (\ref{ineq}).

We will conclude this discussion of the spectrum by providing an
alternative form of the expression (\ref{4.7}) which holds for gauge
invariant states. This form will be useful in comparing our result
with those obtained in the loop representation (where, from the
beginning, one restricts oneself to gauge invariant states.) Let
$\Psi_\g$ be a gauge invariant cylindrical function on $\ab$. Then,
the Gauss constraint implies that, at every vertex $v$ of $\g$, the
following condition must hold:
\be\label{4.10}
\sum_{I} X^i_I\,\cdot\,\Psi_\g = 0,
\ee
where $I$ labels the edges of $\g$ at the vertex $v$ and $X^i_I$ is
assigned to the point $v$ and vertex $e_I$ (see (\ref{X}). Therefore,
\be 
\Jd^i\ +\ \Ju^i\ +\ \Jt^i\ =\ 0.  
\ee 
Hence, one can now express the operator (\ref{4.6}) in an alternate
form, 
\be\label{4.11} 
-\triangle_{S,v} = 2(\Jd)^2 + 2(\Ju)^2 -
(\Jt)^2\, .  
\ee
Furthermore, if it happens that $\g$ has no edges which are tangential
to $S$ at $v$, (\ref{4.10}) implies:
\be
-\triangle_{S,v} = 4(\Jd)^2 =  4(\Ju)^2,
\ee
whence the corresponding {\it restricted} eigenvalues of $\hat{A}_S$
are given by $\sum \Pl^2\sqrt{j(j+1)}$, where $j$ are half-integers.
\end{subsection}

\begin{subsection}{Extended spin networks}
\label{s4.2}
%\begin{appendix}
%\renewcommand{\appendix}{\setcounter{equation}{A.}}
%\setcounter{apppendix}{A}

As a prelude to the discussion on eigenvectors, in this sub-section we
will generalize the constructions and results obtained in
\cite{6,7,16} on spin networks and spin network states.  The previous
work showed that the spin network states provide us with a natural
orthogonal decomposition of the Hilbert space $\tilde\Ho$ of gauge
invariant states in to {\it finite} dimensional sub-spaces. Here, we
will extend those results to the space $\Ho$.

We begin by fixing some terminology.  Given $N$ irreducible
representations $\pi_1, ..., \pi_N$ of $SU(2)$, an associated {\it
invariant tensor} ${c^{m_{k+1}...m_N}}_{m_1...m_k}$ is a multi-linear
map from $\bigotimes_{I=1}^{k} \pi_I$ to $\bigotimes_{I=k+1}^{N}
\Pi_I$ such that:
\be
\pi_{k+1}(g)_{m_{k+1}}^{n_{k+1}}\,... \,\pi_N (g)_{m_N}^{n_N} 
{c^{m_{k+1}...m_N}}_{m_1...m_k}\, \pi_1(g^{-1})^{m_1}_{n_1}\, ... 
\, \pi_N(g^{-1})_{m_k}^{n_k}
=\ {c^{n_{k+1}...n_N}}_{n_1...n_k}, 
\ee 
for arbitrary $g\in SU(2)$, where $\pi_I(g)$ is the matrix
representing $g$ in the representation $\pi_I$.  An invariant tensor
${c_{m_1...m_k}}^{m_{k+1}...m_N}$ is also called an {\it intertwining
tensor} from the representations $\pi_{1},...,\pi_k$ into
$\pi_{k+1},...,\pi_N$.  All the invariant tensors are given by the
standard Clebsch-Gordon theory.
    
An {\em extended spin network} is an quintuplet $(\g,\vpi,\c,
\vrho,\M)$ consisting of:

\begin{itemize}
\item[i)] A graph $\gamma$; 
\item[ii)] A labeling $\vpi:=(\pi_1,..,\pi_N)$ 
of the edges $e_1,...,e_N$ of that graph $\gamma$ with irreducible 
and non-trivial representations of 
$SU(2)$;
\item[iii)] A labeling $\vrho:=(\rho_1,...,\rho_V)$ of the vertices
$v_1,...,v_V$ of $\g$ with irreducible representations of $SU(2)$, the
constraint being that for every vertex $v_\alpha$ the representation
$\rho_\alpha$ emerges in the decomposition of the tensor product of
representations assigned by $\vpi$ to the edges intersecting $v_\a$;
\item[iv)] A labeling $\vec{c}=(c_1,...,c_V)$
of the vertices $v_1,...,v_V$ of $\g$ with certain invariant tensors,
namely, assigned to a vertex $v_\alpha$ is an intertwining tensor
$c_\a$ from the representations assigned to the edges coming to $v_\a$
and $\rho_\a$ to the representations assigned to the outgoing edges at
$v_\a$; and,
\item[v)] A labeling $\M := (M_\a)_{\a=1,...,V}=(M_1,...,M_V)$ of the
vertices $v_1,...,v_V$ of $\g$ which assigns to every vertex $v_\a$ a
vector $M_\a$ in the representation $\rho_\a$;
\end{itemize}

It should be emphasized that every $\pi_I$ is necessarily non-trivial
whereas $\rho_\a$ may be trivial (i.e., 1-dimensional).  In the gauge
invariant context \cite{6,7}, $\rho_\a$ are all trivial whence the
items iii) and v) are unnecessary. The details of these conditions may
seem somewhat complicated but they are necessary to achieve the
orthogonal decomposition (\ref{2.5}).

{}From spin networks, we can construct states in $\Ho$. An {\em
extended spin network state} $\N_{\g,{\vec c},{\vec M}}$ is simply a
$C^\infty$ cylindrical function on $\ab$ constructed from an extended
spin network $(\g,\vpi,\vrho,\c,\M)$,
\be
\label{2.1}
\N_{\gamma,\vec{c},{\vec M}}\,(\Ab):=  \big[\bigotimes_{I=1}^N\,\,
\pi_I(\Ab(e_I))\otimes\bigotimes_{\alpha=1}^V M_\alpha\big]\,\cdot\, 
\big[\otimes_{\alpha=1}^V c_\alpha\big],  
\ee 
for all $\Ab \in \ab$, where, as before, $\Ab(e_I)$ is an element of
$G$ associated with an edge $e_I$ and `$\cdot$' stands for
contracting, at each vertex $v_\alpha$ of $\g$, the upper indices of
the matrices corresponding to all the incoming edges, the lower
indices of the matrices assigned to all the outgoing edges and the
upper index of the vector $M_\a$ with all the corresponding indices of
$c_\alpha$.  (We skip $\vpi$ and $\vrho$ in the symbol for the  extended
spin network function because the intertwiners $c$ contain this
information.)  Thus, for example, in the simple case when the network
has only two vertices, and all edges originate at the first vertex and
end at the second, $\N_{\g,\vec{c},\vec{M}}$ can be written out
explicitly as:
\be 
\N_{\g,\vec{c},{\vec M}} = \pi_1(\Ab(e_1))^{n_1}_{m_1}\,\, ... \,\,
\pi_N(\Ab(e_N))^{n_N}_ {m_N}\,\, M_1^{{m'}_1}M_2^{{m'}_2}\,\, 
{c_1^{m_1 ... m_N}}_{{m'}_1}\, c_2{}_{n_1 ... n_N{m'}_2}\, ,
\ee
where indices $m_I,n_I$ range over $1,..., 2j_I+1$ and ${m'}_\a$
ranges over $1, ..., 2j_{\a +1}$. Given any spin network, (\ref{2.1})
provides a function on $\ab$ which is square-integrable with respect
to the measure $\mu^o$. Given an extended spin network function on
$\ab$, the the range $R(\g)$ of the associated graph $\g$ is  
completely determined. Thus, two spin networks can define the same
function on $\ab$ if one can be obtained from the other by
subdividing edges and
changing arbitrarily the orientations. 

It turns out that the spin network states provide a decomposition of
the full Hilbert space $\Ho$ into {\it finite} dimensional orthogonal
sub-spaces (compare with \cite{6,7}).  Given a triplet $(\g, \vpi,
\vec{\rho})$ defined by $(i-iii)$ above, consider the vector space
${\cal H}_{\g, \vpi, \vec{\rho}}$ spanned by the spin network
functions $\N_{\g, \vec{c}, \M}$ given by all the possible choices for
$\vec{c}, \M$ compatible with fixed labelings $\vpi$,
$\vec{\rho}$. Note that, according to the representation theory of
compact groups, every ${\cal H}_{\g, \vpi, \vec{\rho}}$ is a finite
dimensional irreducible representation of $\Gb$ in $\Cyl$. The group
acts there via
\be
\N_{\g, \vec{c}, \M}(g^{-1}\Ab g)\ =\ \N_{\g, \vec{c}, \M'}(\Ab),\ 
\ \ {M'}_\alpha\ =\ \rho_\a(g(v_\a)) M_\a.  
\ee 
Modulo the obvious completions, we have the following orthogonal
decomposition,
\be\label{2.5} 
\Ho\ =\, \bigoplus_{R(\g), \vpi , \vec{\rho}} {\cal H}_{\g, \vpi, 
\vec{\rho}} 
\ee
where, given a graph $\g$, the labelings ${\vpi}$ and $\vec{\rho}$
range over all the data defined above by (i-iii) whereas for $\g$ in
the sum we take exactly one representative from every range of an
analytic graph in $\S$. When $\vrho$ is trivial we skip $\rho$ in
${\cal H}_{\g, \vpi, \vec{\rho}}$. On $ {\cal H}_{\g, \vpi}$, the
action of the gauge transformations group $\Gb$ is trivial and we have
the following orthogonal decomposition of the Hilbert space of gauge
invariant cylindrical functions, 
\be\label{2.6} \tilde\Ho\ =\ \bigoplus_{R(\g), \vpi } 
{\cal H}_{\g, \vpi}, 
\ee 
where we used the same conventions as in (\ref{2.5}). Thus, we recover 
the result on spin network states obtained in \cite{6,7}.

We conclude this sub-section with a general comment on spin network
states. Consider {\it tri-valent} graphs, i.e. graphs $\g$ each vertex
of which has three (or less) edges. In this case, the standard
Clebsch-Gordon theory implies that the the number of associated gauge
invariant spin network functions is severely limited: the
corresponding sub-space of $\tilde\Ho$ is one dimensional. Hence, on
the sub-space $\Cyl$ of $\tilde\Ho$ corresponding only to tri-valent
graphs, the (normalized) spin network states provide a natural
orthonormal basis. What is remarkable is that these spin networks were
first introduced by Penrose \cite{17} already twenty five years ago to
probe the microscopic structure of geometry, although in a different
context. Because of the simplicity (and other attractive properties)
of these Penrose spin network states it is tempting to hope that they
might suffice also in the present approach to quantum gravity. Indeed,
there were conjectures that the higher valent graphs are physically
redundant.  However, it turns out that detailed physical
considerations rule out this possibility; quantum gravity seems to
need graphs with unlimited complexity.
\end{subsection}

\begin{subsection}{Eigenvectors}
\label{s4.3}

We are now ready to exhibit eigenvectors of the operators
$\triangle_{S,v}$ and $\hat{A}_S$ for any of the potential eigenvalues
found in section 4.1. We will begin with the full, non-gauge invariant
Hilbert space $\Ho$ and consider an arbitrary surface $S$. Since $\Ho$
serves as the (gravitational part) of the kinematical Hilbert space in
theories in which gravity is coupled to spinor fields, our
construction is relevant to that case. In the second part of this
sub-section, we will turn to the gauge invariant Hilbert space
$\tilde\Ho$ and exhibit eigenvectors for the restricted range of
eigenvalues presented in section \ref{s4.1}.

Fix a point $v$ in the surface $S$. We will investigate the action of
the operators $(\Jd)^2$, $(\Ju)^2$, $(\Jdu)^2$ and $\triangle_{S,v}$
on extended spin network states.  Without loss of generality we can
restrict ourselves to graphs which are adapted to $S$ and contain $v$
as a vertex, say $v=v_1$.  Given a graph $\g$ and labeling $\vpi$ and
$\vec{\rho}$ of its edges and vertices by representations of $SU(2)$,
we shall denote by ${\cal C}_v$ the linear space of the intertwining
tensors which are compatible with $\vpi$ and $\vec{\rho}$ at $v$ in
the sense of section \ref{s4.2}. Let $(\g,\vec{\pi}, \vec{\rho},
\vec{c}, \vec{M})$ be an extended spin network and $\N_{\g,\vec{c},
\vec{M}}$ be the corresponding state. As one can see from
Eqs. (\ref{4.4}, \ref{4.3}), each of the four operators above is given
by a linear combination (with constant coefficients) of gauge
invariant terms of the form $b_{i_1...i_E}X^{i_1}_{I_1}...
X^{i_E}_{I_E}$ where $b_{i_1...i_E}$ is a constant tensor and all the
$X$s are associated with the point $v$ and the edges which meet
there. On $\N_{\g,\vec{c}, \vec{M}}$ the action of any operator of
this type reduces to a linear operator $o_v$ acting in ${\cal
C}_v$. More precisely, if ${\cal O}$ is any of the above operators, we
have
\be 
{\cal O} \N_{\g,\vec{c}, \vec{M}}\ =\ \N_{\g,\c{}', \vec{M}} 
\ee
where $\N_{\g, \c{}', \vec{M}}$ is again an extended spin network
state and the network $(\g,\vec{\pi}, \vec{\rho}, \vec{c}{}',
\vec{M})$ differs from the first one only in one entry of the labeling
$\c{}'$ corresponding to the vertex $v$; ${c'}_\a = c_\a$ for all the
vertices $v_\a\not= v$ and ${c'}_1\, =\, o_vc_1$. Consequently, the
problem of diagonalizing these operators reduces to that of
diagonalizing a {\it finite} symmetric matrix of $o_v$. Note that a
constant vector $M$ assigned to $v$ does not play any role in this
action and hence will just make eigenvectors degenerate.

In the case of operators $(\Jd)^2, (\Ju)^2$ and $(\Jdu)^2$, the
(simultaneous) eigenstates are given by the group representation
theory.  We can now spell out the general construction.

Let us fix a graph $\g$ and arrange the edges that meet at $v$ into
three classes as before; $e_1,...,e_d;\,e_{d+1},....,e_u;\,e_{u+1},
..., e_t$. Let us also fix a labeling $\pi_1,...,\pi_t$ of these edges
by irreducible, non-trivial representations of $SU(2)$ and an
irreducible (possibly trivial) representation $\rho$ which emerges in
the decomposition of $\pi_1\otimes...\otimes\pi_t$. Consider now the
following ingredients:
\begin{itemize}
\item[a)] Irreducible representations $\mu_{(d)}$, $\mu_{(u)}$ and 
$\mu_{(d+u)}$;
\item[b)] Invariant tensors ${c_{(d)}\,}^{m_1...m_dm'}$,
$c_{(u)\,}^{m_{d+1}...m_um''}$ and ${c_{{(u+d)}\,m'm''}}^m$
associated, respectively, to the representations $\pi_1,...,\pi_d,
\mu_{(d)}$, and to $\pi_{d+1},...,\pi_{u}, \mu_{(u)}$ and finally to
$\mu_{(d)}, \mu_{(u)}, \mu_{(d+u)}$; and,
\item[c)] Invariant tensor ${c_{(t)\,_n}}^{m_{u+1}...m_tm}$ associated to
$\mu_{(d+u)},\pi_{u+1},...,\pi_t,\rho$.
\end{itemize}
{}From this structure, construct the following invariant tensor,
\ba\label{c}
c^{m_1...m_tn}\ :=\ {c_{(d)}\,}^{m_1...m_dm'} 
{c_{(u)}\,}^{m_{d+1}...m_{u}m''}{c_{(d+u)\,m'm''}}^{n}
{c_{(t)\,n}}^{m_{u+1}...m_{t}m},
\ea
associated with the representations $\pi_1,...,\pi_t,\rho$. To obtain
a non-trivial result in the end, we need all the tensors to be
non-zero. The existence of such tensors is equivalent to the following
two conditions on the data (a-c):
\begin{itemize}
\item[d)] The representations $\mu_{(d)}$ and $\mu_{(u)}$ emerge
respectively, in $\pi_1\otimes...\otimes\pi_d$ and
$\pi_{d+1}\otimes...\otimes\pi_u$; and,
\item[e)] the representation $\mu_{(d+u)}$ emerges both in
$\mu_{(d)}\otimes\mu_{(u)}$ and
$\pi_{u+1}\otimes...\otimes\pi_t\otimes\rho$.
\end{itemize}
Finally, introduce an extended spin network $(\g, \vpi,\vrho,\c,\M)$
such that 
\be\label{esn} \vpi=(\pi_1,...,\pi_t,...,\pi_N),\ \
\vrho=(\rho,\rho_2,...,\rho_V),\ \ \c= (c,c_2,...,c_V),  
\ee 
the remaining entries being arbitrary.  Then, the corresponding state
$\N_{\g,\c,\M}$ is an eigenvector of the operators $(\Jd)^2,(\Ju)^2,$
and $(\Jdu)^2$ with the eigenvalues $\jd(\jd+1)$, $\ju(\ju+1)$ and
$\jdu(\jdu+1)$, respectively, where the half integers $\jd,\ju$ and
$\jdu$ correspond to the representations $\mu_{(d)},\mu_{(u)}$ and
$\mu_{(d+u)}$.  Hence, this $\N_{\g,\c,\M}$ is also an eigenvector of
$\triangle_{S,v}$ with the eigenvalue (\ref{4.7},\ref{4.8}). It is
obvious, that for any triple of representations $\mu_{(d)},\mu_{(u)}$
and $\mu_{(d+u)}$ satisfying the constraint (\ref{4.8}) there exists
an extended spin network (\ref{esn}).

This construction provides {\it all} eigenvectors of
$\triangle_{S,v}$.  The key reason behind this completeness is that,
given any choice of $\pi_1,...,\pi_d,..., \pi_u,...,\pi_t$ and $\rho$
as above, the invariant tensors which can be written in the form
(\ref{c}) with any $\mu_{(d)},\mu_{(u)}$, and $\mu_{(d+u)}$ span the
entire space ${\cal C}_v$ of invariant tensors at $v$ compatible with
that data. Since the defining formula for a spin network function
(\ref{2.1}) is linear with respect to every component of $\c$, given
any spin network $(\g,\vpi,\vrho,\c,\M)$ it suffices to decompose the
component $c_1$ of $\c$ at $v_1=v$ into invariant tensors of the form
(\ref{c}) in any manner to obtain a decomposition of the corresponding
spin network function into a linear combination of extended spin
network functions given by (\ref{c},\ref{esn}). The desired result now
follows from the orthogonal decomposition of $\Ho$ in to the extended
spin network subspaces.

Let us now turn to the operator $\hat{A}_S$. A basis of eigen vectors
can be obtained in the following way.  Since the area operator can be
expressed in terms of and commutes with $(\Jd)^2,(\Ju)^2$, and
$(\Jdu)^2$ at any point $v$ in $S$, we can simultaneously diagonalize
all these operators. Because for every graph the area operator
preserves the subspace of spin-network states associated with that
graph and for two different graphs the spin network spaces spaces are
orthogonal, it is enough to look for eigen vectors for an arbitrary
graph $\g$. Given a graph $\g$, labelings $\vpi$, $\vrho$ and $\M$ as
in section 4.2, at every vertex $v$ contained in the surface $S$
choose a basis in the space ${\cal C}_v$ consisting of invariant
tensors of the form (\ref{c}). The set of the spin network functions
(\ref{2.1}) constructed by varying $\g,\vpi,\vrho, \M$ and picking at
each vertex $v$ an element of the basis in ${\cal C}_v$ constitutes a
basis in $\Ho$. (If we restrict the labelings to $\rho$ consisting
only of the trivial representations, then the resulting set of spin
network states provide a basis for the space $\tilde \Ho$ of gauge
invariant functions.) Each of such states is automatically an
eigenvector of ${\hat A}_S$ with eigenvalue (\ref{4.9}).

We conclude the first part of this sub-section with a simple example
of an eigen vector of the area operator with eigenvalue $a_S$, where
$a_S$ is any real number satisfying (\ref{4.9},\ref{4.9}).
\begin{itemize}
\item{Example.}  Suppose $(\jd_\alpha, \ju_\alpha, \jdu_\alpha)$, $\a
=1,...,W$, is a finite set of triples of half integers which for every
$\a$ satisfy (\ref{4.8})). Rather than repeating the construction {\it
a)-e)} above step by step, we will specify only the simplest of the
resulting (extended) spin-networks. In $S$ choose $W$ distinct points
$v_\a$, $\a = 1,...,W$.  To every point $v_\a$ assign two finite
analytic curves $e_{d,\a}$ and $e_{u,\a}$ starting at $v_\a$, not
intersecting $S$ otherwise, and going in opposite directions of
$S$. For a graph $\g$ take the graph $\{e_{d,1}, e_{u,1},...,e_{d,W},
e_{u,W}\}$, the vertices being the intersection points $v_\a$ and the
ends of the edges $e_{d,\a}$ and $e_{u,\a}$ (the curves being chosen
such that the points $v_\a$ are the only intersections).  Label each
edge $e_{d,\a}$ with the irreducible representation $\pi_{d,\a}$
corresponding to a given $\jd_\a$ and every edge $e_{u,\a}$ with the
irreducible representation $\pi_{u,\a}$ defined by $\ju_\a$.  That
defines a labeling $\vec{\pi}$ of $\gamma$.  (The absence of edges
$e_{t,\a}$ is equivalent to introducing these edges in any manner and
assigning to them the trivial representations.) To define a labeling
$\vec{\rho}$ at the vertices $v_\a$, assign to every vertex $v_\a$ a
representation $\rho_\a$ defined by a given $\jdu_\a$.  Next, to each
vertex $v_\a$ assign an invariant tensor $c_\a^{m_dm_um}$ associated
to the triple of representations $(\pi_{d,\a}, \, \pi_{u,\a},\,
\rho_\a)$ introduced above.  The construction of a spin-network is
completed by: $(i)$ labeling that end point of each $e_{d,\a}$ and,
respectively, of $e_{u,\a}$ which is not contained in $S$, with the
representation $\rho_{d,\a}:=\pi_{d,\a}$ and, respectively,
$\rho_{u,\a}:=\pi_{u,\a}$; (ii) labeling of these ends of the edges
with the unique invariants corresponding to the representations
$\mu_{(d),\a}, \rho_{(d),\a}$ or, respectively, to $\mu_{(u),\a}
,\rho_{(u),\a}$; $(iii)$ defining a labeling $\vec{M}$ of vertices
which can be chosen arbitrarily, provided at a vertex $v_\a$ the
associated vector $M_\a$ belongs to the representation
$\rho_{(d+u),\a}$ and at an endpoint of either of the edges
$e_{d/u,\a}$ the associated $M_{d/u,\a}$ belongs to $\rho_{(d/u),\a}$.
\end{itemize}
\medskip

As we noted in section 2, the Hilbert space $\Ho$ is the quantum
analog of the full phase space. Now, in the classical theory, the
imposition of the Gauss constraint on the phase space does not
restrict the allowed values of the functional $A_S$ of (\ref{2.4}). It
is therefore of interest to see if this feature persists in the
quantum theory: Is the spectrum of $\hat{A}_S$ on the full $\Ho$ the
same as that on its gauge invariant sub-space $\tilde\Ho$?  As was
indicated in Sec. 4.1, the answer is in the affirmative only if the
surface is open. If $S$ is closed, there are restrictions on the
spectrum which depend on topological properties of $S$ embedded in
$\S$.  The second part of this section is devoted to this issue.  As
indicated in Sec. 4.1, we need to consider three separate cases.
\medskip

\noindent {\it Case i):  $\partial S\not= \emptyset$ (and
$\partial S\subset \S$)}.

We will modify the spin-network of the above Example in such a way as
to obtain a gauge-invariant eigenstate {\it without} changing the
eigen value of the area operator.  Let $\g$ and the labeling
$\vec{\pi}$ be the ones defined in the Example. To each vertex $v_\a$
assign one more edge $e_{t,\a}$ beginning in $v_\a$ and {\it
contained} in $S$. Label it by the representation $\pi_{t,\a}$
corresponding to a given $\jdu_\a$ at that point. The labeling
$\vec{\rho}$ is now taken to be trivial. To every point $v_\a$ assign,
as in the Example, an invariant tensor $c_\a$ associated now to the
representations $(\pi_{d,\a}, \, \pi_{u,\a},\, \pi_{t,\a})$.  Every
extension of this data to a spin-network will define a spin-network
state which is gauge invariant at each of the points $v_\a$.  Now, we
need to define a closed spin-network which contains all the edges
$e_{d,\a},\, e_{u,\a},\, e_{t,\a}$ and provides an extension for the
labelings already introduced. For this, we use a key property of the
area operator associated to a surface with boundary: vertices which
lie on $\partial S$ do not contribute to the action of the operator.
Therefore, we can simply extend every edge $e_{t,\a}$ within $S$ to
the boundary of $S$. Denote the intersection point with $\partial S$
by $v_{t,\a}$. Next, for every $\a$ we extend (in a piecewise analytic
way) the edges $e_{d,\a}$ and $e_{u,\a}$ such that they end at
$v_{t,\a}$. The extended edges form a graph $\g' =\{{e'}_{d,1},
{e'}_{u,1},{e'}_{t,1}, ..., {e'}_{d,W},{e'}_{u,W},{e'}_{t,W}\}$. Let
us label each primed edge by the irreducible representation assigned
before to the edge it is an extension of. This defines a labeling
$\vec{\pi'}$ of $\g'$. Finally, assign to each new vertex $v_{t,\a}$
the non-zero invariant tensor ${c'}_{t,\a\, m_u,m_d,m_t}$ (which is
unique up to rescaling) associated to the triplet of representations
$(\pi_{d,\a}, \, \pi_{u,\a},\, \pi_{t,\a})$. This completes the
construction of a gauge invariant extension of a spin-network state
constructed in the Example. Thus, for an open surface, the spectrum of
the area operator $\hat{A}_S$ on $\tilde\Ho$ is the same as that on
$\Ho$.
\bigskip\goodbreak

\noindent {\it Case ii):  $\partial S= \emptyset$ and $S$ splits $\S$ 
in to two open sets}.

In this case we can not repeat the above construction: Since $S$ has
no boundary, if additional vertices are needed to close the open
spin-network, they must now lie in $S$ and can make unwanted
contributions to the action of the area operator. Consequently, there
are further restrictions on the possible eigenvalues of the operators
$({\Jd})^2$, $({\Ju})^2$ and $({\Jdu})^2$. To see this explicitly,
consider arbitrary spin-network state $(\g, \vec{\pi}, \vec{c})$ given
by the construction {\it a)-e)} of Sec 4.3.  Let $\{v_1,...,v_W \}$ be
a set of the vertices of $\g$ contained in the surface $S$. Graph
$\gamma$ can be split into three graphs: $\gamma_t$ which is contained
in $S$, $\gamma_u$ which is contained in a one side of $S$ in $\S$ and
$\gamma_d$ contained in the other side of $S$ in $\S$. The only
intersection between the two parts is the set $\{v_1,...,v_W \}$ of
vertices of $\g$ which are contained in $S$.  Let $\g_r$ be one of the
parts of $\g$ (i.e. $r=d$ or $r=u$ or $r=t$). According to the
construction {\it a)-e)}, the labelings $\vec{\pi}$ and $\vec{c}$
define naturally on $\g_r$ an extended spin-network. The labeling of
the edges of $\g_r$ by irreducible representations is defined just by
the restriction of $\vec{\pi}$ to $\g_r$. The labeling of the vertices
by irreducible representations and invariant tensors is defined in the
following way. For the vertices of $\g_r$ which are not contained in
$S$, the labelings are taken to be again the restriction of
$\vec{\rho}$ (which are all trivial) and $\vec{c}$. To a vertex $v_\a$
contained in $S$ we assign the representation corresponding to a given
$j_{r,\a}$ and the invariant tensor $c_r$ defined in $b)$ (for
$r=d,u$) and $c)$ (for $r=t$) of the construction {\it a) - e)}.
Finally, we complete it by arbitrary nonzero labeling $\vec{M}$ of the
vertices with vectors in appropriate representations.  The
construction {\it a) - e)} guarantees that a resulting extended
spin-network state is not zero. Now, for an extended spin-network
$(\g',\vec{\pi}{}', \vec{\rho}{}', \vec{c}{}', \vec{M}{}')$ we have
the following ``fermion conservation law'':
\be\label{fermion}
\sum_v j_{{\rho'}(v)} \ = N \ee
%\ 0  {\rm\ \ modulo\ \ } {\Bbb Z},\ee
%
for some {\it integer} $N$, where $v$ runs through the vertices of a
graph $\g'$ and each $j_{\rho(v)}$ is an half-integer corresponding to
an representation assigned to $v$ by $\vec{\rho}{}'$. In our case we
therefore obtain the restriction:
\be\label{cond}
\sum_\a j_{r,\a}\ = N_r \ee
%\ 0  {\rm\ \ modulo\ \ } {\Bbb Z},\ee
%
for $r=d, u, d+u$ which gives the conditions (\ref{sphere}) listed in
Sec 4.1.  (In fact either two of the above conditions imply the third
one).

The conditions (\ref{cond}) are also sufficient for an eigen vector to
exist. Suppose we are given a set of half integers as in the Example
above, which satisfy the restriction (\ref{cond}).  A statement
``converse to the fermion conservation law'' is that for any set
$\{v_1,...,v_W\}$ of points in $S$ and any assignment $v_\a \ \mapsto\
j_{\a}$ where $j_\a$ are non-negative half integer satisfying
(\ref{cond}), there exists an extended spin-network
$(\g',\vec{\pi}{}', \vec{\rho}{}', \vec{c}{}', \vec{M}{}')$ such that
every $v_\a$ is its index, $j_\a$ corresponds to the representation
assigned to $v_\a$ by $\vec{\rho}{}'$, and for every vertex $v\not=
v_\a$, $\a=1,...,W$, of $\g'$, the representation assigned by
$\vec{\rho}{}'$ is trivial.  {}From extended spin-networks provided by
the above statement it is easy to construct an eigen vector of the
corresponding eigen values.
\bigskip\goodbreak

\noindent {\it Case iii):  $\partial S= \emptyset$ but $S$ does not 
split $\S$.} 

The only difference between this case and the previous one is that now
a graph $\g$ representing an eigen vector is cut by $S$ into two
components: $\g_t$ contained in $S$ and $\g_{d+u}$ which corresponds
to the rest of $\g$. Since $\g_{d+u}$ can be now connected by the
same arguments as above, we prove that a necessary and sufficient
condition for an eigen vector to exists is (\ref{cond}) imposed
only on the half integers $\jdu_\a$.
\end{subsection}
\end{section}

\begin{section}{Discussion}
\label{s5}

In section 1, we began by formulating what we mean by quantization of
geometry: Are there geometrical observables which assume continuous
values on the classical phase space but whose quantum analogs have
discrete spectra?  In the last two sections, we answered this question
in the affirmative in the case of area operators.  In the next paper
in this series we will show that the same is true of other
(`3-dimensional' ) operators. The discreteness came about because, at
the microscopic level, geometry has a distributional character with
1-dimensional excitations. This is the case even in semi-classical
states which approximate classical geometries macroscopically
\cite{9,18}.

We will conclude this paper by examining our results on the area
operators from various angles.

{\sl 1. Inputs}: The picture of quantum geometry that has emerged here
is strikingly different from the one in perturbative, Fock
quantization. Let us begin by recalling the essential ingredients that
led us to the new picture.

This task is made simpler by the fact that the new functional calculus
provides the degree of control necessary to distill the key
assumptions. There are only two essential inputs. The first assumption
is that the Wilson loop variables, $T_\a = {\rm Tr}\, {\cal P} -\exp
\int_\a A$, should serve as the configuration variables of the theory,
i.e., that the Hilbert space of (kinematic) quantum states should
carry a representation of the $C^\star$-algebra generated by the
Wilson loop functionals on the classical configuration space
$\ag$. The second assumption singles out the measure $\tilde\mu^o$.
In essence, if we assume that $\hat{E}^a_i$ be represented by $-i\hbar
{\delta/\delta A_a^i}$, the `reality conditions' lead us to the
measure $\tilde{\mu}^o$ \cite{7}.  Both these assumptions seem natural
from a mathematical physics perspective. However, a deeper
understanding of their {\it physical} meaning is still needed for a
better understanding of the overall situation
\footnote{In particular, in the standard spin-2 Fock representation,
one uses quite a different algebra of configuration variables and uses
the flat background metric to represent it. It then turns out that the
Wilson loops are {\it not} represented by well-defined operators; our
first assumption is violated. One can argue that in a fully
non-perturbative context, one can not mimic the Fock space
strategy. Further work is needed, however, to make this argument
water-tight.}. %

Compactness of $SU(2)$ plays a key role in all our considerations.
Let us therefore briefly recall how this group arose.  As explained in
\cite{12,14}, one can begin with the ADM phase-space in the triad
formulation, i.e., with the fields $(E^a_i,\, K_a^i)$ on $\S$ as the
canonical variables, and then make a canonical transformation to a new
pair $(A_a^i := (\Gamma_a^i +K_a^i), E^a_i)$, where $K_a^i$ is the
extrinsic curvature and $\Gamma_a^i$, the spin-connection of $E^a_i$.
Then, $A_a^i$ is an $SU(2)$ connection, the configuration variable
with which we began our discussion in section 2.  It is true that, in
the Lorentzian signature, it is not straightforward to express the
Hamiltonian constraint in these variables; one has to introduce an
additional step, e.g., a generalized Wick transform \cite{13}.
However, this point is not directly relevant in the discussion of
geometric operators which arise at the {\it kinematical} level. (See,
however, below).  Finally, we could have followed the well-known
strategy \cite{22} of simplifying constraints by using a complex
connection ${}^{\Comp}\!A_a^i := (\Gamma_a^i - i K_a^i)$ in place of
the real $A_a^i$.  The internal group would then have been
complexified $SU(2)$.  However, for {\it real} (Lorentzian) general
relativity, the kinematic states would then have been {\it
holomorphic} functionals of ${}^{\Comp}\!A_a^i$.  To construct this
representation rigorously, certain technical issues still need to be
overcome.  However, as argued in \cite{13}, in broad terms, it is
clear that the results will be equivalent to the ones obtained here
with real connections.

{\sl 2. Kinematics versus Dynamics:} As was emphasized in the main
text, in the classical theory, geometrical observables are defined as
functionals on the {\it full} phase space; these are kinematical
quantities whose definitions are quite insensitive to the precise
nature of dynamics, presence of matter fields, etc.  Thus, in the
connection dynamics description, all one needs is the presence of a
canonically conjugate pair consisting of a connection and a (density
weighted) triad.  Therefore, one would expect the result on the area
operator presented here to be quite robust.  In particular, they
should continue to hold if we bring in matter fields or extend the
theory to super-gravity.

There is, however, a subtle caveat: In field theory, one can not
completely separate kinematics and dynamics.  For instance, in
Minkowskian field theories, the kinematic field algebra typically
admits an infinite number of {\it inequivalent} representations and a
given Hamiltonian may not be meaningful on a given representation.
Therefore, whether the kinematical results obtained in any one
representation actually hold in the physical theory depends on whether
that representation supports the Hamiltonian of the model. In the
present case, therefore, a key question is whether the quantum
constraints of the theory can be imposed meaningfully on $\tilde\Ho$.%
\footnote{Note that this issue arises in {\it any} representation once
a sufficient degree of precision is reached. In geometrodynamics, this
issue is not discussed simply because generally the discussion is
rather formal.} %
Results to date indicate (but do not yet conclusively prove) that this
is likely to be the case for general relativity.  The general
expectation is that this would be the case also for a class of
theories such as super-gravity, which are `near' general relativity.
The results obtained here would continue to be applicable for this
class of theories.

{\sl 3. Dirac Observable:} Note that $\hat{A}_S$ has been defined for
{\it any} surface $S$.  Therefore, these operators will not commute
with constraints; they are not Dirac observables.  To obtain a Dirac
observable, one would have to specify $S$ {\it intrinsically}, using,
for example, matter fields.  In view of the Hamiltonian constraint,
the problem of providing an explicit specification is extremely
difficult.  However, this is true already in the classical theory.  In
spite of this, {\it in practice} we do manage to specify surfaces and
furthermore compute their areas using the standard formula from
Riemannian geometry which is quite insensitive to the details of how
the surface was actually defined.  Similarly, in the quantum theory,
{\it if} we could specify a surface $S$ intrinsically, we could
compute the spectrum of $\hat{A}_S$ using results obtained in this
paper.

{\sl 4. Comparison:} Let us compare our methods and results with those
available in the literature. Area operators were first examined in the
loop representation. The first attempt \cite{9} was largely
exploratory. Thus, although the key ideas were recognized, the very
simplest of loop states were considered and the simplest eigenvalues
were looked at; there was no claim of completeness.  In the present
language, this corresponds to restricting oneself to bi-valent graphs.
In this case, apart from an overall numerical factor (which does,
however, have some conceptual significance) our results reduce to that
of \cite{9}.

A more complete treatment, also in the framework of the loop
representation, was given in \cite{10}. It may appear that our results
are in contradiction with those in \cite{10} on two points. First, the
final result there was that the spectrum of the area operator is given
by $\Pl^2 \sum \sqrt{j_l(j_l+1})$, where $j_l$ are half-integers,
rather than by (\ref{4.9}). However, the reason behind this
discrepancy is rather simple: the possibility that some of the edges
at any given vertex can be tangential to the surface was ignored in
\cite{10}. It follows from our remark at the end of section 4.2 that,
given a surface $S$, if one restricts oneself to only to graphs in
which none of the edges are tangential, our result reduces to that of
\cite{10}. Thus, the eigenvalues reported in \cite{10} do occur in our
spectrum. It is just that the spectrum reported in \cite{10} is
incomplete. Second, it is suggested in\cite{10} that, as a direct
consequence of the diffeomorphism covariance of the theory, local
operators corresponding to volume (and, by implication, area) {\it
elements} would be necessarily ill-defined (which makes it necessary
to by-pass the introduction of volume (and area) elements in the
regularization procedure). This assertion appears to contradict our
finding that the area element $\widehat{\sqrt{g_S}}$ {\it is} a
well-defined operator-valued distribution which can be used to
construct the total area operator $\hat{A}_S$ in the obvious
fashion. We understand \cite{23}, however, that the intention of the
remark in \cite{10} was only to emphasize that the volume (and area)
elements are `genuine' operator-valued distributions; thus there is no
real contradiction.

The difference in the methodology is perhaps deeper. First, as far as
we can tell, in \cite{10} only states corresponding to trivalent
graphs are considered in actual calculations. Thus, even the final
expression (Eq (48) in \cite{10}) of the area operator after the
removal of the regulator is given only on tri-valent
graphs. similarly, their observation that every spin network is an
eigenvector of the area operator holds only in the tri-valent case.
Second, for the limiting procedure which removes the regulator to be
well-defined, there is an implicit assumption on the continuity
properties of loop states (spelled out in detail in \cite{24}). A
careful examination shows that this assumption is {\it not} satisfied
by the states of interest and hence an alternative limiting procedure,
analogous to that discussed in section 3.1, is needed. Work is now
progress to fill this gap \cite{23}. Finally, not only is the level of
precision achieved in the present paper significantly higher but the
approach adopted is also more systematic. In particular, in contrast
to \cite{10}, in the present approach, the Hilbert space structure is
known {\it prior} to the introduction of operators. Hence, we can be
confident that we did not just omit the continuous part of the
spectrum by excising by fiat the corresponding sub-space of the Hilbert
space.

Finally, the main steps in the derivation presented in this paper were
sketched in the Appendix D of \cite{7}. The present discussion is
more detailed and complete.

{\sl 5. Manifold versus Geometry:} In this paper, we began with an
orientable, analytic, 3-manifold $\S$ and this structure survives in
the final description. As noted in footnote 1, we believe that the
assumption of analyticity can be weakened without changing the
qualitative results.  Nonetheless, a smoothness structure of the
underlying manifold will persist. What is quantized is `geometry' and
not smoothness. Now, in 2+1 dimensions, using the loop representation
one can recast the final description in a purely combinatorial fashion
(at least in the so-called `time-like sector' of the theory). In this
description, at a fundamental level, one can avoid all references to
the underlying manifold and work with certain abstract groups which,
later on, turn out to be the homotopy groups of the
`reconstructed/derived' 2-manifold (see, e.g., section 3 in
\cite{2+1}). One might imagine that, if and when our understanding of
knot theory becomes sufficiently mature, one would also be able to get
rid of the underlying manifold in the 3+1 theory and introduce it
later as a secondary/derived concept. At present, however, we are
quite far from achieving this.

In the context of geometry, however, a detailed combinatorial picture
{\it is} emerging. Geometrical quantities are being computed by
counting; integrals for areas and volumes are being reduced to genuine
sums. (However, the sums are {\it not} the `obvious' ones, often used
in approaches that {\it begin} by postulating underlying discrete
structures. In the computation of area, for example, one does not just
count the number of intersections; there are precise and rather
intricate algebraic factors that depend on the representations of
$SU(2)$ associated with the edges at each intersection.) It is
striking to note that, in the same address \cite{Rie} in which Riemann
first raised the possibility that geometry of space may be a physical
entity, he also introduced ideas on discrete geometry. The current
program comes surprisingly close to providing us with a concrete
realization of these ideas.

\end{section}

\bigskip\bigskip
{\centerline {\bf Acknowledgments}}

Discussions with John Baez, Bernd Bruegman, Don Marolf, Jose Mourao,
Roger Picken, Thomas Thiemann, Lee Smolin and especially Carlo Rovelli
are gratefully acknowledged. Additional thanks are due to Baez and
Marolf for important information they communicated to JL on symmetric
tensors in the representation theory. This work was supported in part
by the NSF Grants PHY93-96246 and PHY95-14240, the KBN grant 2-P302
11207 and by the Eberly fund of the Pennsylvania State University. JL
thanks the members of the Max Planck Institute for their hospitality.
Both authors acknowledge support from the Erwin Schrodinger
International Institute for Mathematical Sciences, where the final
version of this paper was prepared.

\end{document}